\documentclass[10pt]{llncs}

\usepackage[title]{appendix}
\usepackage{graphicx}
\usepackage{amsmath}
\usepackage{enumerate}
\usepackage{amsmath,times,amsfonts,amssymb}
\usepackage{pgfplots}
\usetikzlibrary{patterns}
\usepackage{float}
\usepackage{multirow}
\usepackage{comment}
\usepackage{setspace}
\usepackage{algorithm, algorithmic}
\usepackage{flushend}

\usepackage{enumitem}
\usepackage{paralist}
\usepackage{cite}

\pgfplotsset{compat=1.14}

\begin{document}

\title{Confidential Boosting with Random Linear Classifiers for Outsourced User-generated Data }

\author{Sagar Sharma, Keke Chen\inst{}}
 \institute{Data Intensive Analysis and Computing (DIAC) Lab, Kno.e.sis Center, \\Wright State University
\email{\{sharma.74,keke.chen\}@wright.edu}}

\maketitle

\begin{abstract}
User-generated data is crucial to predictive modeling in many applications. With a web/mobile/wearable interface, a data owner can continuously record data generated by distributed users and build various predictive models from the data to improve their operations, services, and revenue. Due to the large size and evolving nature of users data, data owners may rely on public cloud service providers (Cloud) for storage and computation scalability. Exposing sensitive user-generated data and advanced analytic models to Cloud raises privacy concerns. We present a confidential learning framework, SecureBoost, for data owners that want to learn predictive models from aggregated user-generated data but offload the storage and computational burden to Cloud without having to worry about protecting the sensitive data. SecureBoost allows users to submit encrypted or randomly masked data to designated Cloud directly. Our framework utilizes random linear classifiers (RLCs) as the base classifiers in the boosting framework to dramatically simplify the design of the proposed confidential boosting protocols, yet still preserve the model quality. A Cryptographic Service Provider (CSP) is used to assist the Cloud's processing, reducing the complexity of the protocol constructions. We present two constructions of SecureBoost: HE+GC and SecSh+GC, using combinations of homomorphic encryption, garbled circuits, and random masking to achieve both security and efficiency. For a boosted model, Cloud learns only the RLCs and the CSP learns only the weights of the RLCs. Finally, the data owner collects the two parts to get the complete model. We conduct extensive experiments to understand the quality of the RLC-based boosting and the cost distribution of the constructions. Our results show that SecureBoost can efficiently learn high-quality boosting models from protected user-generated data.  
\end{abstract}


\vspace{-0.4cm}

\section{Introduction} \label{sec:intro}
\vspace{-0.35cm}

It is a common scenario in which a data owner delivers services such as search engines, movie recommendations, healthcare informatics, and social networking to its subscribing or affiliated users (henceforth referred as users) via web/mobile/wearable applications. By collecting users' activities such as clickthroughs, tweets, reviews, and other information, the data owner accumulates a large amount of user-related data, which are used to build analytic models aimed at improving the quality of related services and operations, and increase revenues. However, due to the ever-growing size of data and associated computation complexities, data owners often rely on easily available public cloud services (Cloud) to outsource storage and computations. 

The reliance on Cloud for the massive collection of user data along with building powerful big data analytic models raise great concerns of user privacy and intellectual property protection. First, the Cloud's infrastructures, if poorly secured, can be compromised by external hackers which damages the data owner's reputation and users' privacy. Recent data breach incidents involved Target, Ashley Madison, and Equifax \cite{mansfield,unger}. Second, the potential threat of unauthorized retrieval, sharing, or misuse of sensitive data by insiders \cite{chen10,duncan12} are difficult to detect and prevent. The data owners have a great responsibility for protecting the confidentiality of the sensitive data collection and analytics in Cloud. Thus, confidential data mining frameworks for outsourced data are highly desirable to data owners. Note that differential privacy does not fully address the problem, as it does not protect intellectual property and not prevent model-inversion attacks \cite{fredrikson14,shokri16} as models are exposed to adversaries. 

Naive applications of the well-known cryptographic primitives such as the fully homomorphic encryption (FHE) scheme \cite{gentry09}, garbled circuits (GC) \cite{yao86}, and secret sharing \cite{demmler15} in building confidential computing frameworks prove too expensive to be practical \cite{lu16stat,niko13}. A few recent studies \cite{niko13,niko13sp,demmler15,mohassel17} started blending multiple cryptographic primitives and adapted to certain privacy architectures to work around the performance bottlenecks. These ``hybrid'' constructions mix different cryptographic primitives to implement the key algorithmic components of a protocol with reasonable overheads. 

While the hybrid approach is promising, it does not fundamentally address the basic complexity of building a confidential version of a learning algorithm. We believe it is more critical to modify the original algorithm or adopt a ``crypto-friendly" alternative algorithm to significantly reduce the associated complexity. However, the current solutions are mostly focusing on translating the original algorithms to confidential ones, from simple linear algorithms such as linear classifiers and linear regressions \cite{graepel12,niko13sp, mohassel17} with weak prediction power, to powerful yet enormously expensive models such as shallow neural networks \cite{mohassel17}. 

\vspace{-0.35cm}
\subsection{Scope of Work and Contributions}\label{subsec:scope}
\vspace{-0.25cm}
While deep learning methods \cite{lecun15} have dominated the image and sequence-based learning tasks, boosting is among the most powerful methods such as SVM and Random Forest \cite{caruana06} for other prediction tasks. For example, it has also been a popular method (e.g., XGBoost \cite{xgboost}) in learning to rank \cite{chapelle11} and a top choice of many Kaggle competition winners. Surprisingly, no work has sufficiently explored the power of boosting in confidential learning.

The core idea of our SecureBoost approach is to fully utilize the powerful boosting theory \cite{freund99} that requires \emph{only} weak classifiers (e.g., each classifier's accuracy is only slightly exceeding $50\%$ for two-class problems) to derive a powerful prediction model. This flexibility allows us to revise the original boosting algorithm (i.e., AdaBoost \cite{freund99}) that use non-crypto-friendly decision stumps to adopt crypto-friendly \emph{random linear classifiers} as the base classifiers. We consider our work as the first step towards developing confidential versions for other boosting algorithms such as gradient-boosting \cite{friedman01}. 

In the popular AdaBoost framework for classification \cite{freund99}, decision stumps (DS) have been used as the weak classifiers for their simplicity and fast convergence of boosting. Although the training algorithm for a decision stump is quite simple, it is expensive to implement its confidential version due to the associated complexity of secure comparisons. Our core design of confidential boosting is to use random linear classifiers (RLCs) as the weak classifiers. For a linear classifier $f(x) = w^Tx$, where $x$ is the feature vector and $w$ is the parameter vector to learn, an RLC sets $w$ to be random using a specific generation method independent of training data. This random generation of the classifier dramatically simplifies the training step and it only requires to determine whether the random classifier is a valid weak classifier (e.g., accuracy $>50\%$). In experiments, we found that our random RLC generation method works satisfactorily - for every 1-2 random tries we can find a valid weak classifier. The resulting boosting models are comparable to those generated by using decision stumps as base classifiers, although it converges slower. The use of RLC also allows us to conveniently protect feature vectors and labels and to greatly reduce the costs of other related steps. 

We have designed two secure constructions to implement the RLC-based boosting framework to understand the effect of different cryptographic primitives on the associated complexities and expenses. The constructions are based on the non-colluding honest-but-curious Cloud-CSP setting that has been used by recent related work \cite{niko13sp,niko13,mohassel17}. CSP is a cryptographic service provider that will be responsible to manage encryption keys and assist Cloud with the intermediate steps of the boosting framework. Cloud takes over the major computation and storage burden but is not interested in protecting user privacy. Both of our protocols result in models with distributed parameters between the Cloud and the CSP: the Cloud holding the RLCs' parameters and the CSP holding the base classifier's weights of the boosted models.  An alternate setting (i.e., our SecSh setting) is that two servers take an equal share of computation and storage. For simplicity, we unify the two settings to Cloud-CSP.

We carefully analyze the security of the constructions, based on the universally composable (UC) security paradigm \cite{canetti01, canetti15}, and show that no additional information is leaked except for CSP knowing a leakage function. Both the constructions of SecureBoost expose a leakage function to CSP - the correctness of RLC's prediction on training examples. We analyze the leaked information of the function and show that it is safe to use under our security assumption. 

We summarize the unique contributions as follows:
\begin{compactitem}
\item We propose to use random linear classifiers as a crypto-friendly building block to simplify the implementation of confidential boosting. 
\item We develop two hybrid constructions: HE+GC and SecSh+GC, with the combination of GC, SHE, Secret Sharing, AHE, and random masking to show that the RLC-based boosting can be elegantly implemented. 
\item Our framework provably preserves the confidentiality of users' submitted data, including both feature vectors and their associated labels, and the generated boosting models from both curious Cloud and CSP. 
\item We conduct an extensive experimental evaluation of the two constructions with both synthetic and real datasets to fully understand the costs and associated tradeoffs.
\end{compactitem}
\vspace{-0.35cm}

\section{Preliminary} \label{sec:prelim}
\vspace{-0.35cm}
We use lowercase letters for vectors or scalars; capital letters for matrices and large integers; and single indexed lowercase or capital case letters for vectors.

\textbf {Boosting.}Boosting is an ensemble strategy \cite{hastie01} that generates a high-quality classifier with a linear combination of $\tau$ weak base classifiers (whose prediction power is slightly better than random guessing). Specifically, given training examples $\{(x_i, y_i), i=1\dots n\}$, where $x_i$ are feature vectors and $y_i$ are labels, it learns a model  $H(x) = \sum_{t=1}^\tau\alpha_t h_t(x)$, where $h_t$ is a weak classifier that outputs the prediction $\hat{y}$ for the actual label $y$ and $\alpha_t$ is the learned weight for $h_t$. Algorithm \ref{alg:boost} outlines the boosting algorithm for the two-class problem. The most popular weak classifier has been the decision stump \cite{freund99}, which is merely based on conditions like \emph{if $X_j<v_j$, output 1; otherwise, -1}, where $X_j$ is a certain feature and $X_j < v_j$ is some optimal split that gives the best prediction accuracy among all possible single-feature splits for the training dataset. 

\begin{algorithm}
	\caption{Boosting($T$,  $\tau$)}
	\label{alg:boost}
	\begin{algorithmic}
		\small
		\STATE \textbf{input:} training data samples $T=\{(x_i, y_i), i=1\dots n$, where $x_i \in \mathbb{R}$ and $y_i \in \{1,-1\}\}$, number of base classifiers: $\tau$
			\STATE Initialize the sample weights $\delta_{1i} \leftarrow 1/n$ for $i = 1\dots n$;
			\FOR{ $t$ $\leftarrow$ $1$ \TO $\tau$}
				\STATE learn a weak classifier $h_t(x)$ with sample weights $\delta_{t,i}$,$i=1 \dots n$;
				\FOR{$i$ $\leftarrow$ $1$ \TO $n$}
					\STATE $e_{t,i}= 1$ if $h_t(\delta_{t,i}x_i)==y_i$ else $0$;
				\ENDFOR
				\STATE $error = \sum_{i=1}^n{e_{t,i}}\delta_{t,i}$;
				\STATE $\alpha_t = ln((1-error)/error)$;
				\STATE $\delta_{t+1, i} = \delta_{t, i} \exp(\alpha_i e_{t, i})$ for $i=1\dots n$;
				\STATE $\delta_{t+1} = \delta_{t+1}/|\delta_{t+1}|$;  
			\ENDFOR
		\STATE \textbf{Output:} $H(x) = \sum_{t=1}^{\tau}\alpha_t h_t(x)$
	\end{algorithmic}
\end{algorithm} 

\textbf{Additive Homomorphic Encryption.}\label{subsec:ahe}
For any two integers $\alpha$ and $\beta$, an AHE scheme allows the additive homomorphic operation: $E(\alpha+\beta) = f(E(\alpha), E(\beta))$ where the function $f$ works on encrypted values without decryption. For example, Paillier encryption \cite{paillier99} is one of the most efficient AHE implementations. Conceptually\footnote{Paillier encryption allows more efficient multiplication. .}, with one operand, either $\alpha$ or $\beta$, unencrypted, we can derive the \textit{pseudo-homomorphic} multiplication, e.g., $E(\alpha\beta) = E(\sum_{i=1}^{\beta}\alpha)$. Similarly, we can derive pseudo-homomorphic vector dot-product, matrix-vector multiplication, and matrix-matrix multiplication, as long as one of the operands is in plaintext. 
 
\textbf{RLWE Homomorphic Encryption.}\label{subsec:rlwe}
The RLWE scheme is based on the intractability of the learning-with-error (LWE) problem in certain polynomial rings \cite{BGV12}. It allows both homomorphic addition and multiplication. RLWE allows multiple levels of multiplication with a rising cost. For details, please refer to Brakerski et al. \cite{BGV12}. \emph{Message packing} \cite{BGV12} was invented to pack multiple ciphertexts into one polynomial, greatly reducing the ciphertext size - e.g., we can pack about 600 encrypted values (slots) into one degree-12,000 polynomial. With message packing, vector dot-products and matrix-vector multiplication can be carried out efficiently as shown by\cite{halevi14algo}.

\textbf{Randomized Secret Sharing.}\label{subsec:SecSh}
The randomized secret sharing method \cite{demmler15} protects data by splitting it into two (or multiple) random shares, the sum of which recovers the original data, and distributing them to two (or multiple) parties. Several protocols have been developed to enable fundamental operations such as addition and multiplication based on distributed random shares, producing results that are also random shares, such as the multiplicative triplet generation method \cite{demmler15,mohassel17}. 

\textbf{Garbled Circuits.}\label{subsec:GC}
Garbled Circuits (GC) \cite{yao86} allow two parties, each holding an input to a function, to securely evaluate a function without revealing any information about the input data. The function is implemented with a circuit using a number of basic gates such as AND and XOR gates. The truth table of each gate is encrypted so that no information is leaked during the evaluation. One party creates the circuit and the other one evaluates it. All inputs are securely encoded as labels and passed to the evaluator via the 1-out-of-2 Oblivious Transfer (OT) \cite{asharov13} protocol. During the recent years, a number of optimization techniques have been developed to minimize the cost of GC, such as free XOR gates \cite{kolesnikov08}, half AND gates \cite{zahur15}, and OT extension \cite{asharov13}. 

\vspace{-0.35cm}
\section{Framework}\label{sec:framework}
\begin{figure}[h]	
 \vspace{-0.3cm}
 \centering	
 \includegraphics[width=0.55\linewidth]
 {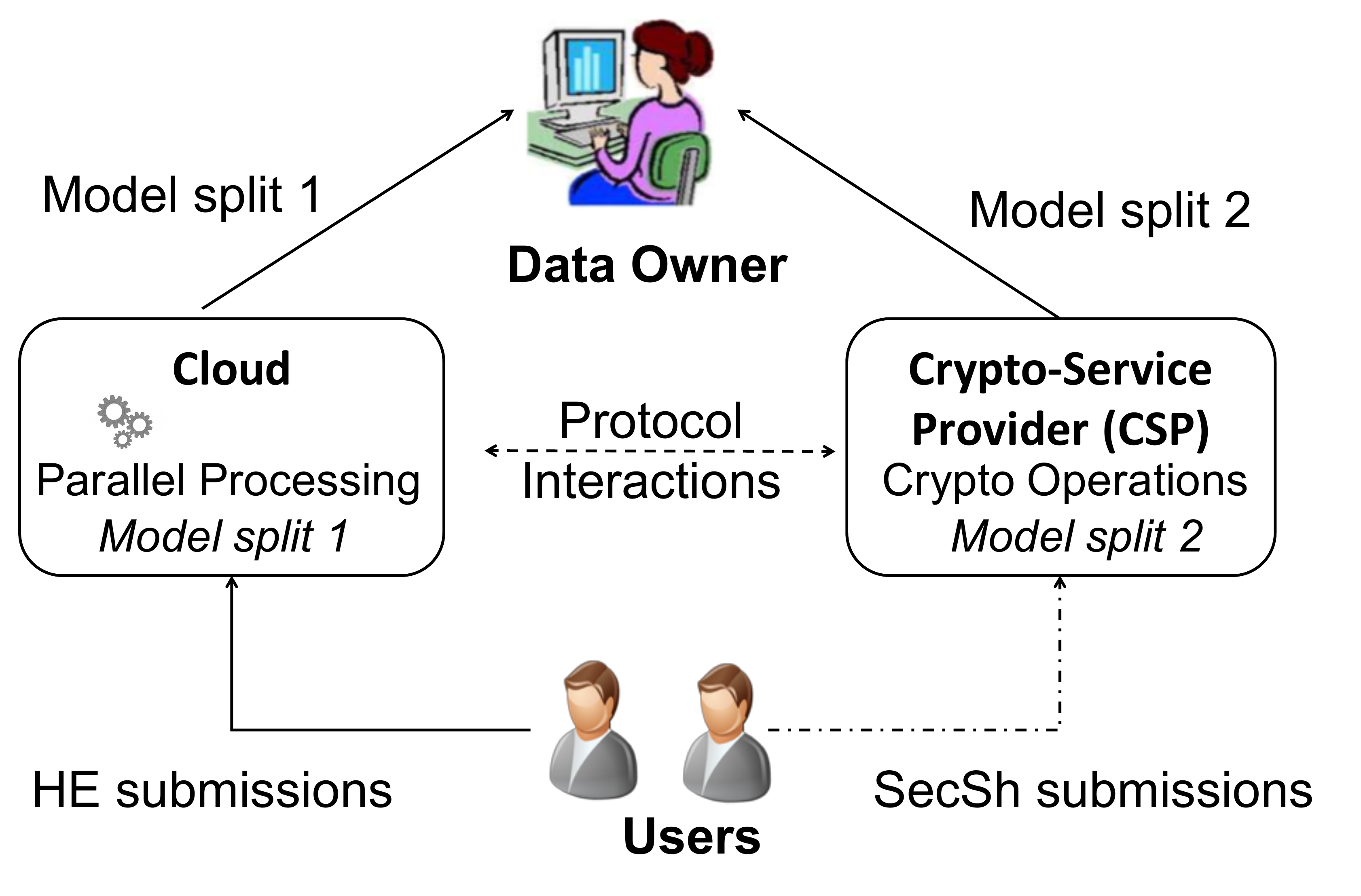}
 \vspace{-0.3cm}
 \caption{SecureBoost Framework. 
}
 \label{fig:framework}	
 \vspace{-0.45cm}
 \end{figure}

\vspace{-0.25cm}
Figure \ref{fig:framework} shows the SecureBoost framework and the involved parties: the data owner, the cloud service provider (Cloud), the users who contribute their personal data for model training, and the Cryptographic Service Provider (CSP). The learning protocol consists of multiple rounds of Cloud-CSP interactions, which builds a boosted model on the global pool of user-contributed training data. Ultimately, Cloud learns the parameter of each base classifier but no additional knowledge about the protected user data; and CSP learns the weights of the base classifiers and a certain type of leakage information that does not help breach the confidentiality of protected user data. The learned models can be either downloaded and reconstructed by the data owner for local applications or used by data owner by submitting encrypted new records to Cloud and undergoing Cloud-CSP evaluation. 

Data owner designates a cloud provider to collect user-generated data in encrypted form and undertake the major storage cost and the major computation-intensive components of the confidential learning protocol. CSP is a party with limited resources. It mainly assists Cloud in intermediate steps, e.g. encrypting or decrypting intermediate results and constructing garbled circuits. CSP is allowed to learn some leakage function but remains oblivious to users' data or the models learned. The concept of CSP has been used and justified by other approaches \cite{niko13,niko13sp} as a practical semi-honest setting to release data owner from complex interactions. If using randomized secret sharing, the users upload shares of their submissions to both Cloud and CSP as depicted by the dotted lines in Figure \ref{fig:framework}.

\vspace{-0.35cm}
\subsection{SecureBoost Learning Protocol} \label{sec:protocol}
\vspace{-0.35cm}
In this section, we describe the rationale and benefits of using RLCs as the base classifiers, the major components of the SecureBoost protocol, and the security goals.

\vspace{-0.35cm}
\subsubsection {RLCs as Base Classifiers}
The original boosting framework has used decision stumps as the base classifiers. RLCs are overly ignored due to its slower convergence rate. However, it is expensive to implement decision stumps on encrypted data due to the $O(kn\log n)$ comparisons in the optimal implementation, where $n$ is the number of records and $k$ is the dimensionality. It is known that comparison on encrypted data is expensive for both homomorphically encrypted data \cite{lu16stat} or garbled circuits \cite{lazzeretti11}. To eliminate the cost of comparison, we use randomly generated linear classifiers (RLC) instead. An RLC generates a classification plane in the form of $h(x) = w^Tx + b$ with randomly selected $w$ and $b$, which can be done by one party, i.e., Cloud. Thus, no comparison is needed in base-classifier generation. 

\begin{figure}[h]	
\vspace{-0.35cm}
 \centering	
 \includegraphics[width=0.55\linewidth]
 {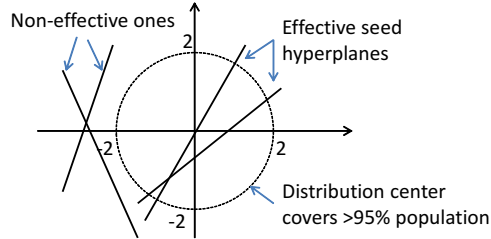}
 \vspace{-0.45cm}
 \caption{Effective Random Linear Classifier Generation}
 \label{fig:rlc_effectiveness}	
 \vspace{-0.55cm}
 \end{figure}

However, blindly selecting $w$ and $b$ is not efficient. As Figure \ref{fig:rlc_effectiveness} shows, the generated plane needs to shatter the training data space into two partitions of significant sizes. For this purpose, we require the submitted data to be normalized so that the training vectors are distributed around the origin. In practice, with the standardization procedure, i.e., each dimension $X_i$ is normalized with $(X_i-\mu_i)/\sigma_i$, where $\mu_i$ is the mean and $\sigma_i$ is the standard deviation of the dimension $X_i$, most dimensional values should be in the range $[-2, 2]$. Thus, we can choose $b$, the intercept, in the range $[-2, 2]$, while each element of $w$ is chosen uniformly from [-1, 1]. Note that $\mu_i$ and $\sigma_i$ can be roughly estimated with low-cost sampling and aggregation protocols from users' submissions. For clarity, we ignore the details of these simple protocols.  With this setting, we find in our experiments that a valid random linear classifier can be found in about 1-2 tries. We have also verified with our experiments that boosting with RLCs can generate high-quality models comparable to those with decision stumps. 

RLCs have extra advantages. First, they allow learning with both the feature vectors and labels protected. We can transform the training data as $x \leftarrow (x, 1)$ and $w\leftarrow (w, b)$, with which the  hypothesis function simply changes to $h(x) = w^Tx$. For a two-class problem with labels $y\in \{-1, 1\}$, if the result $h(x)$ gives a correct prediction, i.e., the same sign as the label $y$, we always get $h(x)y = w^Txy > 0$; otherwise $w^Txy \leq 0$. Note that $xy$ stays together in the evaluation, and thus users can submit the encrypted version of $xy$, $E(xy)$, protecting both feature vectors and labels. Second, they simplify the learning of base classifiers. As $w$ is randomly generated, there is no need for Cloud to consider sample weights during learning. Meanwhile, the learning of the $\alpha_t$ weights can be individually done by CSP. Finally, this process allows only the CSP to learn the weights of base models, and Cloud to learn the base classifiers, preventing either party learning the final model.

\vspace{-0.45cm}
\subsubsection{SecureBoost Protocol} The SecureBoost learning protocol is defined with a 4-tuple: SB-Learning = (\textbf{Setup, BaseApply, ResultEval, Update}). Algorithm \ref{algo:general_prot} depicts the use of these components in the boosting framework. For a boosted model $H(x)=\sum_{t=1}^\tau \alpha_t h_t(x)$, Cloud learns the base models $\{h_t(x)=w_t^Tx, t=1..\tau\}$, and CSP learns the model weights $\{\alpha_t, t=1..\tau\}$. 

 \begin{algorithm}
 	\caption{ SecureBoost Framework}\label{algo:general_prot}
	\begin{algorithmic}[1]
		\small
		\STATE \textbf{($K, E(Z), \{w_i, i=1..p\}, \delta_1$)$\leftarrow$Setup($1^k$, $\tau$, $p$)};
		\FOR{ $t$ $\leftarrow$ $1$ \TO $p$}
			\STATE \textbf{$\{E(h_t(x_i)), i=1..n\} \leftarrow $ BaseApply($K$, $E(Z)$, $w_t$)};
			\STATE \textbf{$I_t \leftarrow $ ResultEval($K$, $\{E(h_t(x_i), i=1..n\})$};
			\STATE \textbf{($\delta_{t+1}, \alpha_t, e_t$)$\leftarrow$ Update($K$, $\delta_{t}$, $I_t$)}; //by CSP only
			\IF{$\tau$ effective base models have been found}
			\STATE stop the iteration;
			\ENDIF 
		\ENDFOR
	\end{algorithmic}
\end{algorithm}

\textbf{($K, E(Z), \{w_i, i=1..p\}, \delta_1)\leftarrow$Setup($1^k$, $\tau$, $p$):} 
(1) The key $K$ is generated by a certain party or parties (CSP, Cloud, or both) as required, with the desired security level $1^k$ and all public keys are published. (2) CSP initializes $\delta_1$ with $1/n$. (3) The training data $Z$ of $n$ instances contains row vectors $z_i=x_iy_i$, which is protected with either a  public-key encryption scheme or random masking (e.g., in the secret-sharing construction) to generate $E(Z)$. (4) Data owner sets the desired number of classifiers, $\tau$, and instructs Cloud to generate a pool of prospective RLCs with parameters $w_t$ for $t = 1 \dots p$, where $p$ is the pool size proportionally larger than $\tau$, e.g., $p=1.5\tau$. 

\textbf{$\{E(h_t(x_i)), i=1..n\} \leftarrow$ BaseApply($K$, $E(Z)$, $w_t$)}: With the encrypted training data $E(Z)$ and a model parameter $w_t$, the procedure will output the model $h_t$'s encrypted prediction results on all training instances.  

\textbf{$I_t \leftarrow $ ResultEval($K$, $\{E(h_t(x_i)), i=1..n\}$): } With the encrypted prediction results, ResultEval allows CSP (not Cloud) to learn the indicator vector $I_t$ of length $n$, indicating the correctness of $h_t$'s prediction for each training instance. 

\textbf{$(\delta_{t+1}, \alpha_t, e_t)\leftarrow$ Update($\delta_{t}$, $I_t$)}: CSP takes $I_t$, $\delta_t$ to compute the weighted error rate $e_t=I_t^T\delta_t$ and if $h_t$ is a valid base classifier i.e. accuracy $>50\%$, updates its weight $\alpha_t = 0.5 ln((1-e_t)/e_t)$ and computes $\delta_{t+1}$ for the next iteration with sample weight updating formula. 

In the end, Cloud learns $\{w_t, t=1..p\}$ and CSP learns $\{\alpha_t, t=1..p\}$. A two-party function evaluation protocol can be easily developed for Cloud to apply the model for classification, which, however, is not the focus of this paper. The data owner can simply download the model components from the two parties and reconstruct the final model for local application. The design of leaking $I_t$ represents a careful balance between security and efficiency. While it is possible to hide $I_t$, the complexity of Cloud and CSP processing will be dramatically increased. We have carefully studied the implication of $I_t$ in Section \ref{sec:privacy_analysis} and found its impact on security is minimal.

\vspace{-0.45cm}
\subsection{Security Model}\label{subsec:security-model}
\vspace{-0.35cm}
We make some relevant security assumptions here:  (1) Both Cloud and CSP are honest-but-curious parties, i.e., they follow the protocols exactly and provide services as expected. However, they are interested in the users' data. (2) Cloud and CSP do not collude, (3) The data owner owns data and models thus is a fully trusted party, (4) All infrastructures and communication channels are secure. While the integrity of data and computation is equally important, we consider it orthogonal to our study. We are mainly concerned with the confidentiality of the following assets.

\begin{compactitem}
	\item \textbf{Confidentiality of training data.} User-generated training data may include personal sensitive information. We consider both feature values and the labels sensitive. For example, a user's fitness activity dataset may contain sensitive features such as heart rate and locations, while the labels, i.e., the type of activity, may imply their activity patterns and health conditions. 
	\item \textbf{Confidentiality of prediction models.} The learned models are proprietary to the data owner and can link to confidential users' data. Therefore, the model parameters are split and distributed between Cloud and CSP. No single party can learn the complete model. 
\end{compactitem}

We adopt the universally composable (UC) security \cite{canetti01, canetti15} to formally define the protocol security. We consider an ideal protocol $\pi$ implementing the \emph{ideal functionality} $\mathcal{F}$ corresponding to a SecureBoost protocol, involving Cloud and CSP. In the \underline{$Real$} world, an \emph{honest-but-curious adversary} $\mathcal{A}$ can corrupt any of the parties and gain access to all the inputs and outputs of that party. We say that $\pi$ securely realizes $\mathcal{F}$ (or $\pi$ is UC-secure)  if for any $\mathcal{A}$ in real world there exists an ideal-process simulator $\mathcal{S}$ in ideal world running probabilistic algorithms in polynomial time (i.e., PPT), such that for any environment $\mathcal{Z}$ and inputs $m = (m_\mathcal{Z},m_\mathcal{\mathcal{A}/\mathcal{S}},m_{Cloud/CSP})$,
\vspace{-0.2cm}
\[\begin{aligned}
	&|Pr(Real_{\pi, \mathcal{A}, \mathcal{Z}}(k, z, m) = 1) - Pr(Ideal_{\mathcal{F}, \mathcal{S}, \mathcal{Z}}(k, z, m) =1)|&= negl(k),
\end{aligned}\]
where $negl(k)$ is a negligible function \cite{katz07}. In Section \ref{sec:privacy_analysis}, we  propose two theorems that can be proved to show that SecureBoost protocols are UC-secure.
\vspace{-0.40cm}

\section{Construction with HE and GC}\label{sec:core_alg_rlwe}
\vspace{-0.30cm}
In this section, we present the homomorphic encryption (HE) and GC based construction of SecureBoost. With the HE encrypted data, the \textbf{BaseApply} procedure is essentially the homomorphic operation $E(Z)w_t$ that is allowed by both Paillier \cite{paillier99} and RLWE \cite{BGV12} cryptosystems. We use a garbled-circuit based protocol to allow only CSP to learn the indicator vector $I_t$, without leaking any other information to the parties. In the following, we first describe the construction of the protocol components and then discuss several key technical details.

\textbf{Setup.} CSP generates the HE public and private key and distributes the public key to the users and Cloud. The private key accessible to the data owner when necessary. Users encrypt their submissions. Cloud generates the pool of $p$ prospective weak classifier vectors, $\{w_t, t=1..p\}$. 

\textbf{BaseApply.} With the matrix-vector homomorphic operations enabled by HE, Cloud computes $\{E(u_t) = E(Zw_t), t=1..p\}$. As this step can be done locally by Cloud, Cloud may choose to conduct this work \emph{offline} before the protocol interactions start. 

\textbf{ResultEval.} The problem setting is that Cloud holds $E(u_t)$ and CSP securely identifies the sign of each element of $u_t$, i.e., $Zw_t>0$ implying correct prediction by the RLC, which sets the corresponding element of $I_t$ to 1; otherwise to 0. The sign of element is related to the specific integer encoding, which we will elaborate more. With our encoding scheme, we only need to check a specific bit to determine whether $Zw_t>0$ is true. To satisfy all the security goals, we decide to use a GC protocol for this step that will be discussed in more detail. 

\begin{figure}[h]
\centering
	\vspace{-0.1cm}
	\centering
	\includegraphics[width=0.55\linewidth]{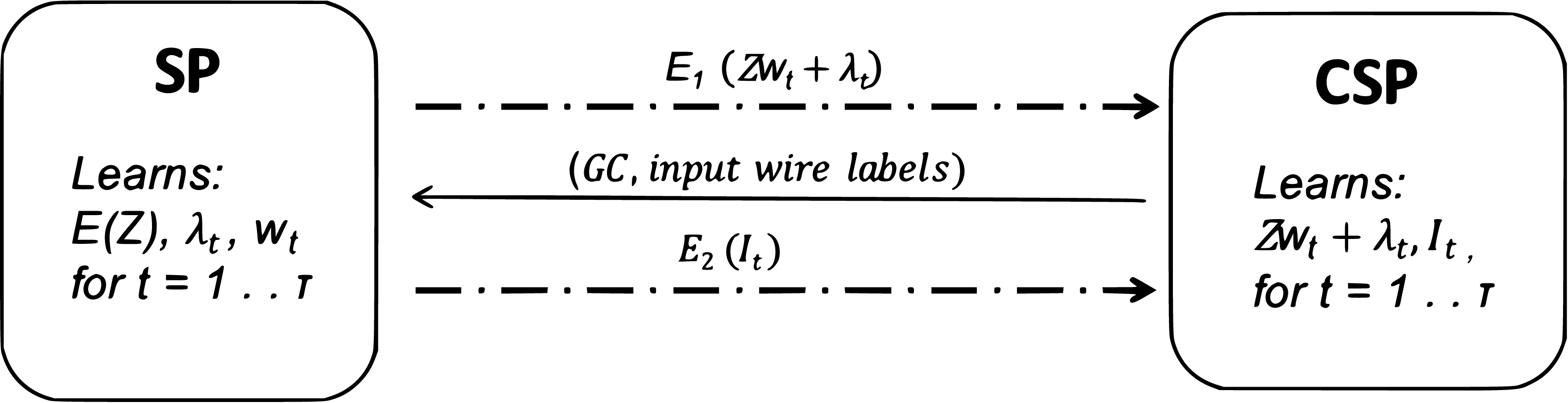}
	\vspace{-0.1cm}
	\caption{Cloud-CSP interactions in HE+GC construction. $E_1$ represents HE encryptions whereas $E_2$ represents GC labels for the GC outputs.}
	\label{fig:messages_HE}
	\vspace{-0.45cm}
\end{figure}

As the last step \textbf{Update} does not involve crypto operations, we can skip its discussion. Algorithm \ref{algo:rlwe_boost} summarizes the operations in this construction and Figure \ref{fig:messages_HE} depicts all the associated Cloud- CSP interactions in this construction. 

\begin{algorithm}
	\caption{HE+GC based SecureBoost}\label{algo:rlwe_boost}
	\begin{algorithmic}[1]
		\small
		\STATE \textbf{Setup}($1^k$, $\tau$, $p$): CSP uses an HE scheme to generate a private and public key pair. CSP initializes $\delta_1$. Users use the public key to encrypt data, and Cloud generates pool of $w_i$;
		\FOR{ $t$ $\leftarrow$ $1$ \TO $p$}
\STATE \textbf{BaseApply($K$, $E(Z)$, $w_t$)}: Cloud computes $E(u_t)= E(Z)w_t$, which can be done offline in batch;
\STATE \textbf{ResultEval($K$, $E(u_t)$)}:  
			\begin{compactitem}
			\item Cloud perturbs $E(u_t)$ with random noise $\lambda_t$, as $E(u_{t,1}) = E(u_t) + E(\lambda_t)$ and sends $E(u_{t, 1})$ to CSP; CSP decrypts it;\\
			\item CSP constructs and Cloud evaluates the garbled circuit that de-noises $u_{t,1}$ as $u_t = u_{t_1} - \lambda$, and returns the indicator vector $I_t$ to CSP\\
			\end{compactitem}
\STATE \textbf{Update($K$, $\delta_{t}$, $I_t$)};
\IF{$\tau$ effective base models have been found}
\STATE stop the iteration;
\ENDIF 
\ENDFOR
	\end{algorithmic}
\end{algorithm}

\vspace{-0.45cm}
\subsection{Technical Detail}\label{subsec:technical_detail}
\vspace{-0.35cm}
Now, we discuss the key problems mentioned in the sketch of the construction above. 

\textbf{Choice of HE Schemes.}
We consider two choices of HE: Paillier \cite{paillier99} and RLWE \cite{BGV12} in our evaluations. 
Paillier scheme provides a large bit space allowing to preserve more precisions in floating-integer conversion. Our evaluation shows that with message packing, all RLWE operations including encryption, decryption, addition and one-level multiplication are much faster than Paillier, although the ciphertext size might be larger than that of Paillier. 

\textbf{Integer Conversion.} 
The HE schemes work on integers only. For a floating-point value $x$, $x\in \mathbb{R}$, to preserve $m$-digit precision after the decimal point upon conversion and recovery, we have: $v = \lfloor 10^mx \rfloor \mod q$, where $q$ is a large integer such that $10^mx \in (-q/2, q/2)$. Let the modulo operation map the values to $[0, q)$, in such a way that the negative values are mapped to the upper range $(q/2, q)$. It is easy to check that $x$ is recoverable: if $v> q/2$, $x\approx (v-q)/10^m$; otherwise, $x\approx v/10^m$.  The modulo additions and multiplications preserve the signs and are thus recoverable. Furthermore, this encoding simplifies the evaluation of the RLC base classifiers, which involves checking the sign of $h_t(x)$. Let $b$ be the total number of bits to represent the values in $[0, q)$. It is trivial to learn that if the $b$-th bit of a value in the range $[0, q)$ is 1, then the value is in the range $(q/2, q)$, which is negative; otherwise, the value is positive. With large enough $q$ we can accommodate the desired multiplication and addition results without overflow. An $n$-bit plaintext space that allows one multiplication followed by $\alpha$ additions, as used in our protocol, spares $(n-\alpha)/2 $ bits to encode the original value. For easier processing, we normalize the original real values in the same dimension of training data before converting them to $b$ bit integers.  

\textbf{Secure Matrix-Vector Multiplication.} 
The core operation $E(Zw_t)$ involves encrypted $E(Z)$ and Cloud generated random plaintext $w_t$. Thus, both AHE and SHE schemes can be applied. 

\textbf{Securely Checking Signs of $E(u_t)$.} 
CSP needs to check the result of base classifier prediction, $E(u_t)=E(Zw_t)$ to learn the correctness of prediction on each instance, so that the error rate, the model weight, and the sample weight update can be computed. With the described integer conversion encoding method, the sign checking $u_{t,i} <0?$ is determined by a specific bit in the result. Note that letting CSP know $u_t$ directly may reveal too much information significantly weakening the security. To balance between security and efficiency, we decide to let CSP only learn the signs indicating if the base classifier $h_t$ correctly classified the training instances, and nothing else is leaked.  Lu et al. \cite{lu16stat} have proposed a comparison protocol based only on RLWE, however, it is extremely expensive to be adapted to our framework. Therefore, we rely on a noise addition procedure to hide the decrypted $u_t$ from CSP and a GC-based de-noising and bit extraction procedure to let CSP learn the specific bit for sign checking. We give the details of these procedures next. 

To hide the plaintext $u_t$ from CSP, we use a noise addition method that can be easily implemented by Cloud on the encrypted vector with homomorphic addition: $E(u_{t,0}) = E(u_t)+E(\lambda_t)$, where $\lambda_t$ is a noise vector generated by the pseudo-random number generator $\mathcal{G}$. Then, CSP can decrypt $E(u_{t,0})$ to learn the noisy result. Let $u_{t,1}=\lambda_t$ held by Cloud. Now the problem is turned to using a GC to securely compute $u_t = u_{t,0} - u_{t,1}$ and return the specific bit of each element of $u_t$. 

\begin{figure}[h!]	
\vspace{-0.3cm}
\centering	
\includegraphics[width=0.4\linewidth]
{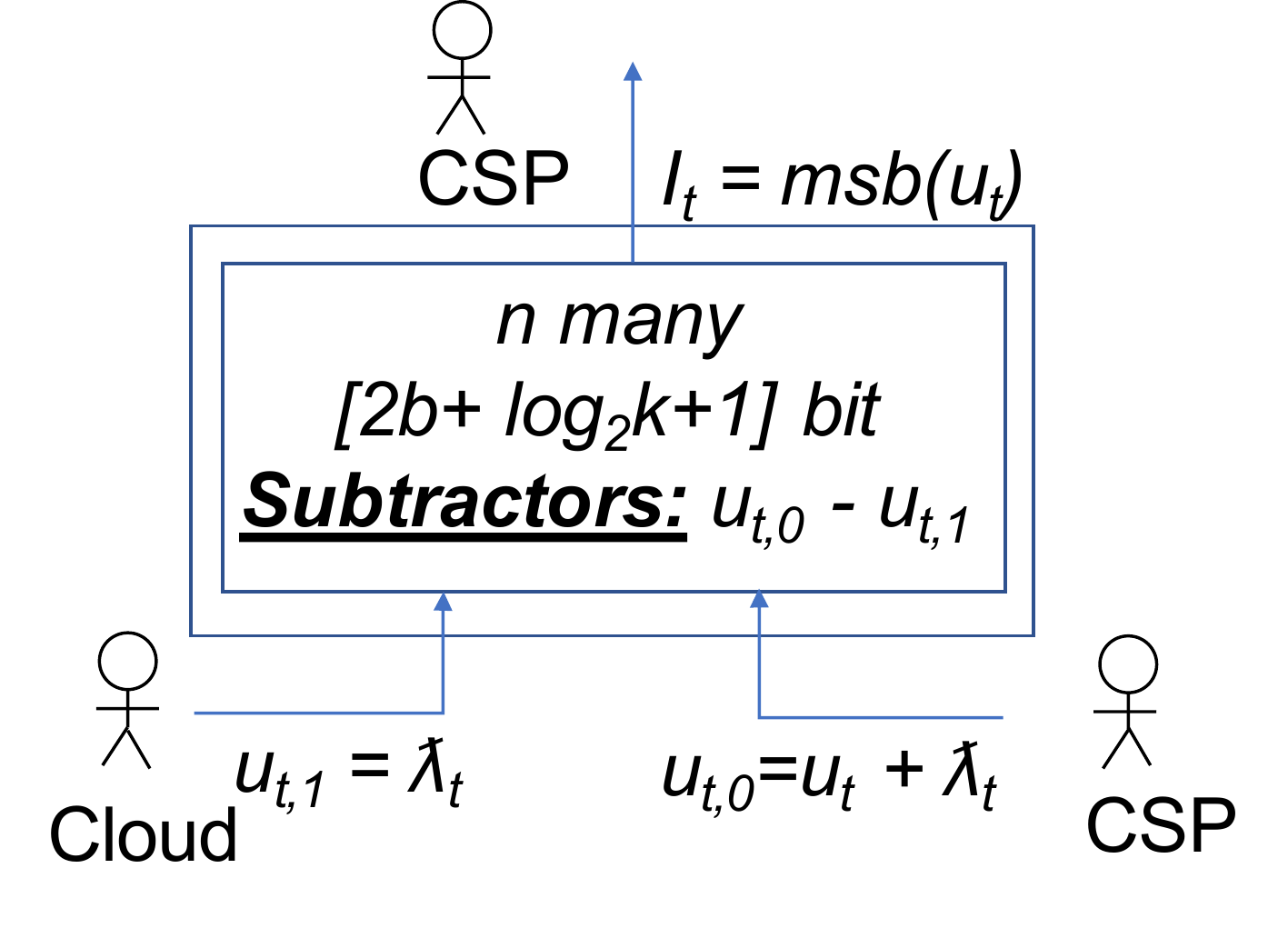}
\caption{GC-based sign checking protocol.}
\label{fig:GC_cmp_import}	
\vspace{-0.3cm}
\end{figure}

Figure \ref{fig:GC_cmp_import} shows the GC based de-noising and bit extraction protocol. CSP's input to the circuit is the binary form of $u_t'$ elements whereas Cloud's inputs are the binary form of $\lambda_t$ elements. With associated oblivious transfer (OT) protocol and wire label transfers, the circuit can securely evaluate $u_t'- \lambda_t$ and extract the most significant bit, $msb(u_{t,j}), j=1..n$, of the result without leaking anything else. Cloud evaluates the circuits and returns the extracted encrypted bits (represented as output labels in GC) to CSP. CSP can then decrypt (re-map) the labels to generate the indicator vector $I_t$.

\vspace{-.35cm}
\section{Construction with {SecSh} and GC}\label{sec:core_alg_SecShare}
\vspace{-0.35cm}
Alternatively, we design our framework with a mixture of secret sharing and garbled circuit techniques. We call this construction ``SecSh + GC''. A somewhat similar approach was taken by \cite{mohassel17} in constructing confidential gradient-descent based learning. It differs from the HE based construction in two aspects: 1) user data protection uses secret sharing, and 2) matrix-vector multiplication happen over secret random splits of training data held by Cloud and CSP.

Instead of encryption, users randomly split their training data into two shares, one for Cloud and the other for CSP. The sum of shares recovers the original values. Any intermediate results that need protection are also in the form of random shares distributed between Cloud and CSP. As a result, multiplication of two values, say, $a$ and $b$, each as random shares (e.g., Cloud holds $a_0$ and $b_0$ while CSP holds $a_1$ and $b_1$, where $a_0+a_1=a$ and $b_0+b_1=b$), needs the help of AHE encryption to compute each party's random share for $ab$. As for sign checking, we reuse the GC protocol designed earlier for HE+GC. 

\textbf{Setup.} Each user splits their data $Z$ into a random matrix $Z_0$ and $Z_1$, where $Z_1 = Z-Z_0$, and securely distributes $Z_0$ to Cloud and $Z_1$ to CSP. Cloud also generates a key pair for a chosen AHE scheme and shares the public key with CSP.

\textbf{BaseApply.} With Cloud holding $Z_0$ and $w_t$, and CSP holding $Z_1$, BaseApply will generate random shares of the result $u_t=Zw_t=u_{t,0} - u_{t,1}$: $u_{t,0}$  and $u_{t,1}$ held by Cloud and CSP, respectively. This is implemented with a special matrix-vector multiplication algorithm, which we will describe later. 

\textbf{ResultEval.} With the random shares: $u_{t,0}$  and $u_{t,1}$ held by Cloud and CSP respectively, we can apply the same GC protocol presented in the last section for computing $u=u_{t,0}-u_{t,1}$ and extracting the specific bits. 

\begin{figure}[h]
	\centering
	\vspace{-0.2cm}
	\includegraphics[width=0.55\linewidth]{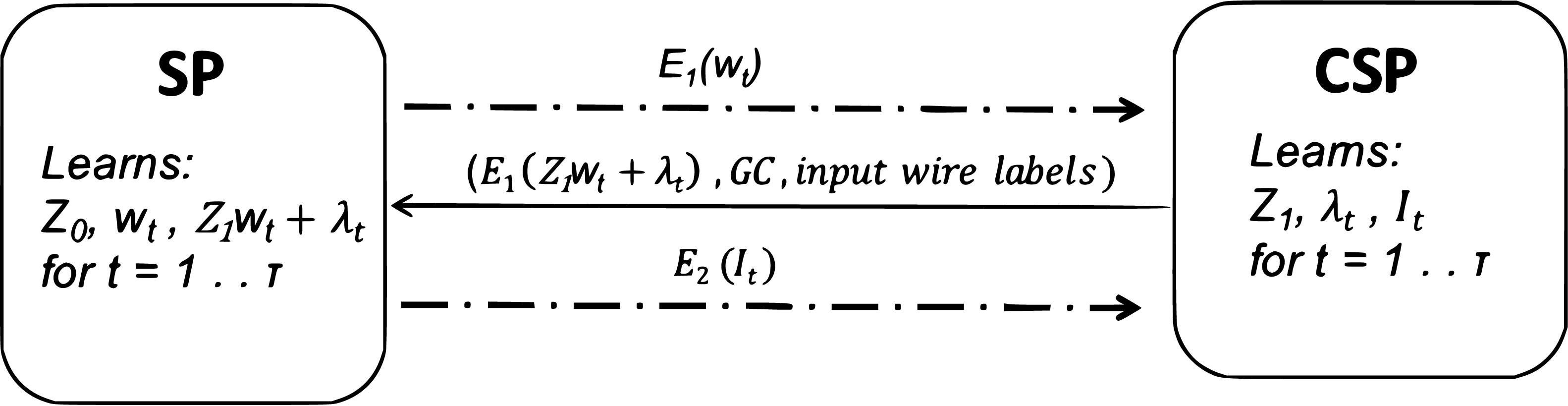}
	\vspace{-0.1cm}
	\caption{Cloud-CSP interactions in the SecSh+GC construction. $E_1$ represents HE encryptions whereas $E_2$ represents GC labels for the GC outputs.}
	\label{fig:messages_sec_sh}
\end{figure}

Algorithm \ref{algo:privacy_SecShare} outlines the construction and Figure \ref{fig:messages_sec_sh} summarizes all the interactions between Cloud and CSP.  

\begin{algorithm}
	\caption{SecSh+GC based SecureBoost}\label{algo:privacy_SecShare}
	\begin{algorithmic}[1]
		\small
		\STATE \textbf{Setup:} Cloud and CSP possess their share of training data $Z_0$ and $Z_1$ respectively; Cloud generates  $p> \tau$ many random vectors $w_t$; Cloud also generates an AHE key pair, and distributes the public key to CSP; CSP initializes the sample weights $\delta_{1,i} = 1/n$ for $i = 1  \dots n$
		\FOR{ $t$ $\leftarrow$ $1$ \TO $p$}
			\STATE \textbf{BaseApply:} apply the random-share based matrix-vector multiplication; in the end, Cloud holds $u_{t, 0}$ and CSP holds $u_{t,1}$
			\STATE \textbf{ResultEval:} use the GC protocol to compute the subtraction of $u_{t,0}$ and $u_{t,1}$ and extract the specific bits of $u_t = u_{t_0} - u_{t_1}$.
\STATE \textbf{Update($K$, $\delta_{t}$, $I_t$)};
\IF{$\tau$ effective base models have been found}
\STATE stop the iteration;
\ENDIF 
		\ENDFOR
		\end{algorithmic}
\end{algorithm}

\vspace{-0.45cm}
\subsection{Technical Detail}
The SecSh+GC construction reuses the integer conversion and the GC-based sign checking components. Here, we focus on the major difference: the protocol for computing matrix-vector multiplication with random shares. 

\textbf{Random-Share-Based Matrix-vector Multiplication.} To initiate, Cloud and CSP respectively hold the two shares $Z_0$ and $Z_1$ of user data in plaintext, and Cloud also holds $w_t$ in plaintext. The goal is to derive random shares of $Zw_t$ and each party learns only one of the shares. 

Cloud computes the part $Z_0w_t$ in plaintext by itself. The challenge is to collect the other part $Z_1w_t$ without CSP knowing $w_t$ and no party knowing the complete result, $Zw_t$. We use the following procedure to achieve this security goal. (1) Cloud encrypts $w_t$ with an AHE scheme and sends $E(w_t)$ to CSP so that CSP can apply pseudo-homomorphic multiplication to compute $E(Z_1w_t)=Z_1E(w_t)$. (2) CSP generates a random vector $\lambda_t$ with the pseudo-random number generator $\mathcal{G}$, encrypts it with the public key provided by Cloud, and apply homomorphic addition to get $E(Z_1w_t + \lambda_t)$, which is sent back to Cloud. (3) Cloud decrypts it and sums up with the other part $Z_0w_t$ to get $Zw_t + \lambda_t$. In the end, Cloud gets $u_{t,0}=Zw_t + \lambda_t$ and CSP gets $u_{t,1}=\lambda_t$. At this point, Cloud and CSP use the GC protocol for sign checking in Section \ref{sec:core_alg_rlwe}. 

\vspace{-0.45cm}
\section{Cost Analysis}\label{sec:cost_analysis}
\vspace{-0.35cm}
Table \ref{tab:big} summarizes the associated big-O estimation of communication and computation broken down into different operations/components. The notations are the same as defined. In summary, we observe that HE+GC constructions demand no CSP storage and CSP only needs to conduct decryptions and GC constructions. In contrast, in SecSh+GC, the workload and storage are almost equally distributed to Cloud and CSP. However, as user-generated data is not encrypted but split into random shares in SecSh+GC, users' costs and overall storage costs are much lower. 

\begin{table*}[t]
\centering
\scriptsize
\caption{BigO estimation for SecureBoost constructions} \label{tab:big}
	\vspace*{-\baselineskip}
	 \resizebox{0.63\columnwidth}{!}{%
	\scriptsize
	\begin{tabular}{|c|c|c|c|c|c|c|c|c|c|}
	\hline
	Construction&Party & Encryption & Decryption& Enc. Mult/Add & Enc. Comm. & GC Comm. & Storage \\
	\hline
	\multirow{3}{*}{HE+GC }&User&$O(nk)$ &-&-& $O(nk)$&-&-\\
	&Cloud & $O(pn)$ & - & $O(pnk)$ & $O(pn)$ & $O(pnb)$ & $O(nk)$ \\
	&CSP & -& $O(pn)$ &-&-&-&- \\
	\hline
	\multirow{3}{*}{SecSh+GC}&User&-&-&-&-&-&-\\
	&Cloud&$O(pk)$ & $O(pn)$  & - &$O(p(n+k))$ &$O(pnb)$& $O(nk)$\\
	&CSP&$O(pn)$ & -  & $O(pnk)$&- &-& $O(nk)$\\
	\hline
	\end{tabular}
	}
	\vspace{-0.45cm}
\end{table*}

\vspace{-0.45cm}
\section{Security Analysis}\label{sec:privacy_analysis}
\vspace{-0.25cm}
According to the security model outlined in Section \ref{subsec:security-model}, we focus on the subcomponents of the protocols that involve both Cloud and CSP and implement a specific ideal function $\mathcal{F}$. The security is proved by finding a simulator $\mathcal{S}$ in the ideal scenario corresponding to the adversary $\mathcal{A}$ in the real scenario such that the environment $\mathcal{Z}$ cannot distinguish the probabilistic outputs of \underline{$Ideal$} and \underline{$Real$}. 

The major interaction happens in computing the indicator vector $I_t$ for an iteration $t$. The corresponding ideal function is defined as $\mathcal{F} (m_{Cloud, t}, m_{CSP, t}) \rightarrow I_t$, where $ m_{Cloud, t}, m_{CSP, t}$ are Cloud's and CSP's inputs to the function and the function's output is the indicator vector $I_t$ as defined by our protocols. We present two theorems next along with their sketch proofs.

\begin{theorem} 
If the random number generator $\mathcal{G}$ is pseudo-random, and both the HE scheme and GC are CPA-secure, then the HE+GC construction of SecureBoost is secure in computing $I_t$ with an honest-but-curious adversary.
\end{theorem}

\textbf{Sketch Proof for Theorem 1.} In the real protocol, the environment machine $\mathcal{Z}$ can observe the inputs $i_{Cloud, t}=(E(Z), w_t, GC_t, Label_{CSP, t})$ and $i_{CSP, t} = (E(Zw_t+\lambda_t), Label_{output, t},   GC_{key})$, and outputs: $o_{Cloud, t}=(\lambda_t, Label_{output, t})$ and $o_{CSP, t} = (GC_t, I_t)$, where $Label$ refers to the GC wire labels. By the security definition of HE, GC, and pseudo-randomness, the variables except for the plaintext $w_t$ and $I_t$ cannot be distinguished from uniformly random values in polynomial time by the environment $\mathcal{Z}$. 

Now, let us define an ideal world parallel to this real protocol. We design an ideal functionality $\mathcal{F}$: with the inputs $m_{Cloud, t}=\lambda_t$ and $m_{CSP, t} = E(Zw_t+\lambda_t)$, it decrypts the inputs, removes the noise $\lambda_t$, and computes the indicator vector $I_t$. Assuming $\mathcal{A}$ compromises Cloud in the real world, we can think of a simulator $\mathcal{S}$ in the ideal world as follows. $\mathcal{S}$ generates a random vector $\lambda_t$ using $\mathcal{G}$ and computes $E(Zw_t+\lambda_t)$ which is sent to CSP and sends $\lambda_t$ to the ideal functionality. Again,  if $\mathcal{G}$ is pseudo-random and HE is CPA-secure, it is impossible for $\mathcal{Z}$ to distinguish $\lambda_t$ and $ E(Zw_t+\lambda_t)$ from uniformly random values in polynomial time. Therefore, both worlds compute the same $I_t$, but $\mathcal{Z}$ cannot distinguish the outputs $\underline{Ideal}$ and $\underline{Real}$ computationally from what it has observed. For a compromised CSP, $S$ generates a random vector $\lambda_t$ using $\mathcal{G}$ and passes it to the ideal functionality. Again, to $\mathcal{Z}$ $\lambda_t$ is indistinguishable from $Zw_t+\lambda_t$. This proves that our protocol $\pi$ \textit{UC-realizes} the ideal functionality $\mathcal{F} (m_{Cloud, t}, m_{CSP, t}) \rightarrow I_t$.

\begin{theorem}
If the random number generator $\mathcal{G}$ is pseudo-random and both the AHE scheme and GC are CPA-secure, then the SecSH+GC construction is secure in computing $I_t$ with an honest-but-curious adversary.
\end{theorem}

\textbf{Sketch Proof for Theorem 2. } In the real protocol, the environment machine $\mathcal{Z}$ can observe the inputs $i_{Cloud, t}=(Z_0, w_t, E_1(Zw_t +\lambda_t), GC_t, Label_{CSP, t})$ and $i_{CSP, t} = (Z_1, E_1(w_t),Label_{output,t}, GC_{key})$, and outputs: $o_{Cloud, t}=(Label_{output, t})$ and $o_{CSP, t} = (\lambda_t, GC_t, I_t)$, where $Label$ refers to the GC wire labels. By the security definition of AHE, GC, and pseudo-randomness, the variables except for the plaintext $w_t$ and $I_t$ cannot be distinguished from uniformly random values in polynomial time by the environment $\mathcal{Z}$. 

Now, let us define an ideal world parallel to this real protocol. We design an ideal functionality $\mathbf{F}$: with the inputs $m_{Cloud, t}= Zw_t+\lambda_t$ and $m_{CSP, t} = E(\lambda_t)$, it decrypts and removes the noise $\lambda_t$, and computes the indicator vector $I_t$. Assuming $\mathcal{A}$ compromises Cloud in the real world, we can think of a simulator $\mathcal{S}$ in the ideal world as follows. $\mathcal{S}$ generates a random vector $\lambda_t$ using $\mathcal{G}$ and computes $(Zw_t+\lambda_t)$ which is passed on to the ideal functionality. Again,  if $\mathcal{G}$ is pseudo-random and AHE is CPA-secure, it is impossible for $\mathcal{Z}$ to distinguish $\lambda_t$ and $ E(Zw_t+\lambda_t)$ from uniformly random values in polynomial time. Therefore, both worlds compute $I_t$, but $\mathcal{Z}$ cannot distinguish the outputs $\underline{Ideal}$ and $\underline{Real}$ computationally from what it has observed. For a compromised CSP, $S$ does the same operation, i.e. generate a random vector $\lambda_t$ using $\mathcal{G}$ and computes $E(Z_1w_t+\lambda_t)$ which is sent to Cloud,  and simply forwards $\lambda_t$ to the ideal functionality. 
This proves that our protocol $\pi$ \textit{UC-realizes} the ideal functionality $\mathcal{F} (m_{Cloud, t}, m_{CSP, t}) \rightarrow I_t$.

\subsection{Implication of Revealing $I_t$ to CSP.}
\vspace{-0.15cm}
CSP learns the indicator function $I_{t,i}(h_t(x_i)==y_i)$, for $i=1..n$ in the iteration $t$ of SecureBoost. It is clear that this leakage does not help CSP learn the complete boosted model $H(x)$ as long as Cloud holds $\{w_t, t=1..\tau\}$ as secrets. However, we must understand if such leakage may help CSP learn anything about the training data. 

Recall that an element of indicator vector  $I_t(h_t(x_i)==y_i)$ represent if the base RLC $h_t$ classifies the training instance $x_i$ correctly or incorrectly (1 and 0, respectively). At the end of learning, each record $x_i$ gets $p$ prediction results for $p$ base classifiers $h_t, t=1..p$, respectively, which is denoted as $c_i = (c_{i,1},\ldots, c_{i,p})$, $c_{i,j}\in \{0, 1\}$.  Let $c_i$ be the \emph{characterization vector} (CV) for the record $x_i$. The intuition tells that two similar records (i.e., relatively small Euclidean distance) with the same label will lead to similar CVs. However, the reverse is uncertain --- if the reverse is true then adversaries can utilize CV similarity to infer record similarity. However, our experiments show that the reverse is clearly false (Figure \ref{fig:char_dist} in Section \ref{sec:experim}). In particular, the records having identical CVs have distances (and their standard deviations) not significantly different from those having other types of CVs.    

\vspace{-0.25cm}
\section{Experiments}\label{sec:experim}
\vspace{-0.25cm}

We design our experiment set on both real and synthetic datasets with three goals: (1) show random linear classifiers are effective weak classifiers for boosting; (2) evaluate associated computation, communication, and storage costs, and their distributions amongst the users, Cloud, and CSP for both the constructions; and (3) understand the trade-off between costs and model quality, including a comparison with another state-of-the-art confidential classification learning framework. 

\textbf{Implementation.} We adopt the HELib library \cite{halevi14algo} for the RLWE encryption scheme, implement the Paillier cryptosystem \cite{paillier99} for the AHE encryption scheme, and use the ObliVM (oblivm.com) library for the garbled circuits.  ObliVM has included the state-of-the-art GC optimization techniques such as half AND gates, free XOR gates, and OT extention. The core algorithms for data encoding, encryption, matrix-vector multiplications, and additive perturbation are implemented with C++ using the GMP library. Users' submissions are encoded with the $7$-bit floating-integer conversion method (Section \ref{subsec:technical_detail}). We use the scikit-learn toolkit (scikit-learn.org) to evaluate the model quality for existing classifier learning methods selected for comparison purpose.  

\textbf{Parameter selection.} We pick cryptographic parameters corresponding to $112$-bit security. The RLWE parameters allow 32-bit message-space overall, $1$ full vector replication, and at least $2$ levels of multiplication. The degree of the corresponding cyclotomic polynomial is set to $\phi(m) = 12,000$ and $c=7$ modulus switching matrices, which gives us $h=600$ slots for message packing. The Paillier cryptosystem uses 2048-bit key-size to achieve approximately 112-bit security. Our GC-based sign checking protocol accommodates $(2b + \log_2(k))$-bit inputs, where $b$ is the bit-precision (i.e., b=7 in experiments) and $k$ is the dimension of the training data.  
Note that HELib uses a text format to store the ciphertext which we zip to minimize the costs.

\textbf{Datasets.}  We test SecureBoost with both the synthetic and real datasets. Table \ref{tab:datasets} summarizes the dataset properties. Datasets are selected to cover a disparate range of dimensions and number of instances.  All selected datasets contain only two classes to simplify the evaluation. The real datasets come from the UCI Machine Learning Repository \cite{frank10}. The synthetic dataset is deliberately designed to generate non-linearly separable classes. It is used to conveniently explore and understand the behaviors of RLC-based boosting and the quality of non-linear classification modeling methods.  
 		
 \begin{table}[h!]
  	\centering
	\vspace{-0.53cm}
  	\scriptsize
  	\caption{Dataset statistics. } \label{tab:datasets}
	 \vspace*{-\baselineskip}
	 \resizebox{0.53\columnwidth}{!}{%
	\tabcolsep=0.11cm
  	\begin{tabular}{|c|c|c|c|c|}
  		\hline
  		Dataset & Instances & Attributes & Adaboost Accuracy & Number of decision stumps \\
  		\hline
		ionosphere & 351&34& 92.02\% +/- 4.26\%&50\\
  		credit & 1,000& 24 & 74.80\% +/- 3.50 \%&100\\
  		spambase& 4,601 & 57&92.31 \% +- 4.40 \%&75 \\
		epileptic&11,500 & 179 & 86.95 \% +- 3.40 \%&200\\
  		synthetic & 150,000 & 10& 89.51 \% +-2.10 \% &75\\
  		\hline
  	\end{tabular}
	\vspace{-0.85cm}
	}
  \end{table} 

\vspace{-0.99cm}
\subsection{Effectiveness of RLC Boosting} 
\vspace{-0.25cm}
The performance of boosting is characterized by the convergence rate and the final accuracy. The speed of convergence is directly related to the overall cost of the SecureBoost protocols. We look at the number of base classifiers ($\tau$) needed to attain a certain level of accuracy. As a randomly generated RLC may fail (i.e., RLCs having $\approx 50\%$ accuracy for the two-class datasets) and be discarded in some of the rounds, we also assess the actual number ($p$) of RLCs that are tried to generate the final model. All the accuracy results are for 10-fold cross-validation. The following results can be reproduced and verified with the scripts we have uploaded to \textit{https://sites.google.com/site/testsboost/}.

\begin{figure}[h]
\vspace{-0.65cm}
\centering
\begin{tabular}{c@{\thinspace}c@{\thinspace}c@{\thinspace}c}
\centering
\begin{tikzpicture}[scale=0.32]
 	\pgfplotsset{every axis legend/.append style={font=\small},every node near coord/.append style={font=\huge}}
 	\begin{axis}
 	[xmode=log,
	xlabel={Number of Base Classifiers $\tau$},xlabel style = {font = \huge},
 	point meta ={y*100},
 	ylabel={Avg. Accuracy}, ylabel style = {font=\huge},
	yticklabel=\pgfmathprintnumber\tick\%,yticklabel={\pgfmathparse{\tick*100}\pgfmathprintnumber{\pgfmathresult}\%},
	legend columns=3,legend style={at={(0.5,0.95)},draw=none,anchor=south,
 		font=\Large},
	y tick label style = {font = \huge},
	x tick label style = {font = \huge}
	]	
	\addplot+[mark=*,error bars/.cd,
 	x dir=both
 	,y dir=both,y explicit]
 	table[x=Base,y=Accuracy,col sep=tab]
 	{./data/ionosphere_ab_cv.csv};
 	\addlegendentry{ionosphere}
 	\addplot+[mark=+,error bars/.cd,
 	x dir=both
 	,y dir=both,y explicit]
 	table[x=Base,y=Accuracy,col sep=tab]
 	{./data/credit_ab_cv.csv};
 	\addlegendentry{credit}	
 	\addplot+[mark=x,error bars/.cd,
 	x dir=both,
 	y dir=both,y explicit]
 	table[x=Base,y=Accuracy,col sep=tab]
 	{./data/spambase_ab_cv.csv};
 	\addlegendentry{spambase}
 	\addplot+[mark=o,error bars/.cd,
 	x dir=both,
 	y dir=both,y explicit]
 	table[x=Base,y=Accuracy,col sep=tab]
 	{./data/epilipsis_ab_cv.csv};
 	\addlegendentry{epileptic}
	\addplot+[mark=square,error bars/.cd,
 	x dir=both,
 	y dir=both,y explicit]
 	table[x=Base,y=Accuracy,col sep=tab]
 	{./data/synthetic_ab_cv.csv};
 	\addlegendentry{synthetic}
	\addplot+[mark=.]
 	table[x=Base,y=Accuracy,col sep=tab]
 	{./data/vertical.csv};
 	\end{axis}
 	\end{tikzpicture}&
	\hspace{-0.4cm}
\begin{tikzpicture}[scale=0.32]
 	\pgfplotsset{every axis legend/.append style={font=\small},every node near coord/.append style={font=\huge}}
 	\begin{axis}
 	[xmode=log,
	xlabel={Number of Base Classifiers $\tau$}, xlabel style = {font=\huge},
 	point meta ={y*100},
 	ylabel={Avg. Accuracy}, ylabel style={font=\huge, yshift = 0.02cm},
	yticklabel=\pgfmathprintnumber\tick\%,yticklabel={\pgfmathparse{\tick*100}\pgfmathprintnumber{\pgfmathresult}\%},
	y tick label style = {font = \huge},
	x tick label style = {font = \huge},
	legend columns=1,legend style={at={(0.65,0.15)},anchor=south, font=\Large,draw=none}]
 	\addplot+[mark=x,error bars/.cd,
 	x dir=both,
 	y dir=both,y explicit]
 	table[x=Base,y=Accuracy,col sep=tab]
 	{./data/synthetic_adaboost_cv.csv};
 	\addlegendentry{Boosting w. DS}
	\addplot+[mark=+,error bars/.cd,
 	x dir=both
 	,y dir=both,y explicit]
 	table[x=Base,y=Accuracy,col sep=tab]
 	{./data/synthetic_lmc_cv.csv};
 	\addlegendentry{Boosting w. LMC}
 	\addplot+[mark=oplus,error bars/.cd,
 	x dir=both
 	,y dir=both,y explicit]
 	table[x=Base,y=Accuracy,col sep=tab]
 	{./data/synthetic_ab_cv.csv};
 	\addlegendentry{Boosting w. RLC}		
 	\end{axis}
 	\end{tikzpicture}&
	\hspace{-0.35cm}
	\begin{tikzpicture}[scale=0.30]
 	\begin{axis}[
 	ybar,x=42,
 	bar width=0.25cm, tick align=inside,
 	ymin=0, ymax=100,
 	axis x line*=bottom,axis y line*=left,
 	y axis line style={opacity=50},
 	tickwidth=3pt,
 	legend columns=2,legend style={at={(0.5,0.99)},draw=none, anchor=south,font=\Large},
 	ylabel={Avg. Accuracy},
	xlabel={Datasets}, xlabel style = {font = \huge},
 	ylabel style={font=\huge, yshift=0.1cm},
 	symbolic x coords={ionosphere,credit,spambase,epileptic,synthetic},
 	xtick=data,
 	yticklabel={\pgfmathparse{\tick}\pgfmathprintnumber{\pgfmathresult}\%},
	y tick label style = {font = \huge},
	x tick label style = {font = \large,rotate=20}]
 	\addplot[error bars/.cd, y dir=both,y explicit][draw=black, fill=red,fill opacity=0.4,postaction={pattern= dots}] coordinates {
		(ionosphere,92.02)+-(4.26,4.26)
		(credit,74.80)+-(3.50,3.50) 
 		(spambase,92.3)+-(4.40,4.40)
		(epileptic,86.95)+-(3.40,3.40)
 		(synthetic,89.51)+-(2.15,2.15)};
 	\addplot[error bars/.cd, y dir=both,y explicit][draw=black, fill=blue,fill opacity=0.85,postaction={pattern= crosshatch}] coordinates {
		(ionosphere,91.45)+-(3.10,3.10)
 		(credit,73.4)+-(2.40,2.40) 
 		(spambase,88.39)+-(4.3,4.30)
		(epileptic,84.97)+-(3.91,3.91)
 		(synthetic,87.59)+-(3.18,3.18) };
 	\legend{Boosting w. DS,Boosting w. RLC}
 	\end{axis}
 	\end{tikzpicture}&
	\hspace{-0.30cm}
 \begin{tikzpicture}[scale=0.32]
 	\begin{axis}
 	[
 	style={font=\large},
 	enlargelimits=0.15,every tick label/.append style={font=\Large},
 	legend columns=3,legend style={at={(0.5,0.89)},draw=none,anchor=south,
 		font=\large},
 	ylabel={Avg. Accuracy}, ylabel style = {font=\huge},
	xlabel={Precision bits (b)}, xlabel style = {font = \huge},
 	symbolic x coords={5,7,9,11,15,20,real},
 	x tick label style={font=\huge,rotate=20},	yticklabel={\pgfmathparse{\tick}\pgfmathprintnumber{\pgfmathresult}\%},
	y tick label style ={font = \huge}
	]
	\addplot +[mark=*,error bars/.cd, y dir =both, y explicit]
		coordinates{(5,88.6) (7, 89.7) (9,89.5) (11, 90.6) (15,87.5 ) (20, 88.1) (real,91.5)};
	\addplot +[mark = +,error bars/.cd, y dir =both, y explicit]
		coordinates{(5,69.708) (7, 74.546) (9,74.296 ) (11, 74.444) (15,74.472 )(20, 74.252)(real,75.201)};
	\addplot +[mark=x, error bars/.cd, y dir =both, y explicit]
		coordinates{(5,89.1) (7, 88.5)(9,89.7)(11,89.1 )(15,89.3)(20,87.7)(real,91.16)};
	\addplot +[mark=o, error bars/.cd, y dir =both, y explicit]
		coordinates{(5,83.7) (7, 84.5)(9,84.6)(11,83.8 )(15,84.4)(20,85.7)(real,84.97)};
	\addplot +[mark=square,error bars/.cd, y dir =both, y explicit]
		coordinates{(5,84.20) (7, 85.32) (9,85.44) (11, 85.22)(15,84.43)(20, 85.29)(real,92.00)};
	\legend{ionosphere,credit, spambase, epileptic,synthetic}
 	\end{axis}x
 	\end{tikzpicture}\\
	\vspace{-0.2cm}
	\small (a) & \small (b)&\small (c)&\small (d)
	\end{tabular}
	\vspace{-0.4cm}
	\caption{(a) Convergence of boosting with RLCs. (b) Convergence of boosting with RLCs, LMCs, and DSes for the synthetic dataset. (c) Model quality: boosting with RLCs vs. boosting with DSes. (d) Bit precision vs. model accuracy}
 	\label{fig:convergence_cmp}
	\vspace{-0.65cm}
\end{figure}
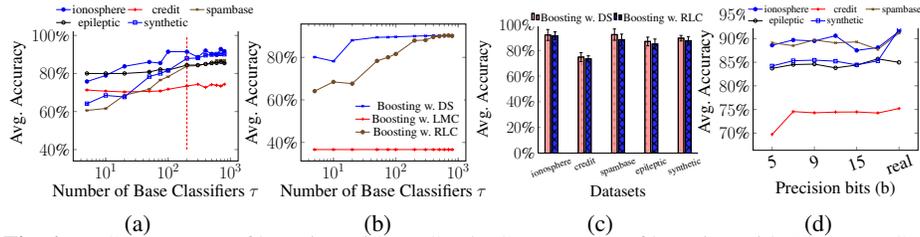

Figure \ref{fig:convergence_cmp}(a) analyzes the convergence of RLC-based boosting for each dataset. We observe that overall only about 200 base classifiers are sufficient to reach a stable model accuracy level for the considered datasets. Figure \ref{fig:convergence_cmp}(b) compares boosting with different base classifiers: RLC, decision stumps (DS), and linear means classifiers (LMC) when learning on the synthetic dataset. Clearly, DS has the advantage of converging faster in about 75-80 rounds. On the other hand, boosting with LMC does not reach the desired accuracy, because the centers of class (i.e., the ``means'') that are used to define the classification plane stay stable even with changed sample weights. The result is a bunch of highly similar base classifiers in the final boosting model, which does not take advantage of the boosting framework. 

Figure \ref{fig:convergence_cmp}(c) shows the final model quality produced by RLC boosting and the DS boosting (i.e., the default boosting method). We use 200 RLCs and varying number of DSes as shown in Table \ref{tab:datasets} as the base classifiers for the datasets. In every case, both methods generate models with almost identical accuracy. All of the above results suggest that RLC boosting is robust and generates high-quality classification models.  
 
\textbf{Encoding Bits.}
The number of bits for encoding affects the cost of GC-related components and the precision in floating-integer conversion, which in turn affects the final model quality. Figure \ref{fig:convergence_cmp} (d) shows the effect of preserved bits on model accuracy. It seems preserving 7 bits is sufficient to get optimal quality models. 
	
\textbf{Cost comparison with DS.} As there is no DS learning algorithm on encrypted data (possibly due to its high expense), we develop a DS learning protocol that fits our framework to estimate the costs as shown in Appendix \ref{App:DS_boost}.  

\vspace{-0.45cm}
\subsection{Cost Distribution} \label{subsec:cost_experim} 
\vspace{-0.25cm}
We now inspect the associated costs for each involved party in the two constructions. Table \ref{tab:setting} shows the parameter settings for different datasets that led to the desired model quality. $\tau$ is the number of base classifiers in the final boosting model. $p$ represents the total number of RLCs that are tried in the modeling process, which determines the actual protocol costs. Overall, in about 1-2 tries on average, we can find a valid RLC (with accuracy $>50\%$).

\begin{table}[h]
 	\vspace{-0.55cm}
  	\centering
  	\scriptsize
  	\caption{Parameter setting for cost evaluation. $\tau$ and $p$ - number of desired and tried RLCs} \label{tab:setting}
	 \vspace*{-\baselineskip}
	  \resizebox{0.30\columnwidth}{!}{%

  		\begin{tabular}{|c|c|c|c|c|c|}
  			\hline
  			Dataset & $\tau$ & $p$ & Accuracy\\
  			\hline
			ionosphere & 200 & 226 &91.5\% +/- 3.1\% \\
  			credit & 200 & 342 & 73.4 \% +/- 2.4 \% \\
  			spambase & 200 & 229 & 87.4 \% +/- 4.8 \% \\
			epileptic& 200 & 331 & 84.41\% +/-2.9 \% \\
			synthetic& 200 & 244 & 87.91\% +/-3.2 \% \\
  			\hline
  		\end{tabular}
		}
		\vspace{-0.55cm}
  	\end{table} 

\textbf{User's Costs.}
A user's costs depend on the size of training data, i.e. the number of training records $n$, and the number of dimensions $k$ per record. The Paillier+GC construction requires each user to encrypt their submission element-wise in streaming or batched manner. The RLWE+GC construction requires each user to batch her submissions and encrypt them as a column-wise matrix $E(Z)$ with message packing (see Section \ref{subsec:rlwe}).  For the SecSh+GC construction, users simply apply the one-time padding method to generate the masks and distribute the splits to Cloud and CSP, respectively. 
\begin{table}[h]
  	\centering
	\vspace{-0.55cm}
  	\scriptsize
  	\caption{User's cost for a batch of 600 records} \label{tab:datacontributor}
	\vspace*{-\baselineskip}
	 \resizebox{0.37\columnwidth}{!}{%
  	\begin{tabular}{|c|c|c||c|}
  		\hline
		\multirow{2}{*}{Dataset}&\multicolumn{2}{c||}{HE+GC (RLWE / Paillier)} & \multicolumn{1}{c|}{SecSh+GC}\\
  		& Enc. (secs) & Upload (MB)&Upload. (MB) \\
  		\hline
		ionosphere&1.54 / 235.83 & 38.50 / 10.25 &0.04\\
  		credit &1.09 / 168.45 &27.50/ 7.32&0.03\\
  		spambase& 2.54 / 390.80&63.80/ 16.99&0.07 \\
		epileptic& 7.91 / 1,212.84 &198.0 / 52.73 &0.09 \\
  		synthetic& 0.48 / 74.12 & 12.1 / 3.22 &0.05 \\
  		\hline
  	\end{tabular}
	}
	\vspace{-0.55cm}
  \end{table} 
  
Table \ref{tab:datacontributor} depicts the user's costs in encrypting and submitting \emph{one batch of records} with the batch size $h=600$.  The HE+GC constructions are more expensive than SecSh+GC in all aspects, but still quite acceptable in most cases.  RLWE+GC results in larger ciphertext but far less computations than Paillier+GC. 

\textbf{Cloud and CSP Costs.}
As Cloud's and CSP's costs are highly inter-related in the SecureBoost constructions we discuss them together. Note: We use the Paillier cryptosystem in SecSh+GC as the required AHE scheme. Table \ref{tab:cost_dist} sums up the costs for all the components. For the smaller datasets, the RLWE+GC construction does not show much benefit over the other two. For datasets with the larger number of records such as the synthetic dataset, both Cloud and CSP take less computational time with RLWE+GC construction in comparison with the other two. For datasets with larger dimensions such as the epileptic dataset, RLWE+GC is more onerous to the Cloud whereas beneficial to the CSP in terms of computation cost. As for storage and communication costs, Paillier+GC and SecSH+GC are favorable across the board. 

 \begin{table*}[h] 	
 	\vspace{-0.45cm}
 	\centering
	\scriptsize
 	 \caption{Overall Cloud and CSP Costs: Storage, Comp. (computation), Comm.(communication)}  \label{tab:cost_dist}
	 \vspace*{-\baselineskip}
	 	\resizebox{0.8\columnwidth}{!}{%
 	 	\begin{tabular}{|c|||c||c|c||c|||c|c||c|c||c|}
 		\hline
 		\multirow{3}{*}{Dataset}& \multicolumn{4}{c|||}{HE+GC (RLWE / Paillier)} & \multicolumn{5}{c|}{SecSh+GC} \\ 
		\hline
		&\multicolumn{1}{c||}{Storage(MB) } & \multicolumn{2}{c||}{Comp. (minutes) }& Comm. (MB) & \multicolumn{2}{c||}{St.(MB)} & \multicolumn{2}{c||}{Comp. (minutes)} & Comm.(MB) \\
		&Cloud &Cloud&CSP&&Cloud&CSP&Cloud&CSP&\\
 		\hline
		ionosphere &38.5 / 6.0 & 13.5 / 21.1 & 3.5 / 16.3 & 286.2 / 81.0 &2.6 &2.6& 17.8 & 19.6 &84.8  \\
 		credit & 55.0 / 12.2 & 28.0 / 83.2 & 12.9 / 70.5 & 1,119.2 / 537.2 & 8.1 & 8.1& 72.1 & 81.6& 541.3  \\
 		spambase & 510.4 / 130.3 & 129.5 / 358.6 & 33.3 / 268.6 & 3,842.6 / 1,876.6 & 76.4 & 76.4& 271.8 & 355.3 & 1,885.1 \\
 		epileptic & 3,960.0 / 1,010.7 & 932.2 / 1,453.0 & 128.2 / 777.0 & 12,291.6 / 6,868.3 & 653.4 &653.4& 788.1 & 1,441.8 & 6897.4 \\
		synthetic &3,025.0 / 805.7 & 1,414.7 / 8,147.3 & 1,175.4 / 7,424.0 &106,891.1 / 57,662.2& 383.9& 383.9& 7,424.5 & 8,146.8 & 57,663.5  \\	
 		\hline
		\end{tabular}%
		}
	\vspace{-0.55cm}
 \end{table*}
 
Next, we analyze the shared GC components for the selected real and synthetic datasets in Table \ref{tab:datasets}. As all the constructions share the same GC component for sign checking, we list the GC costs together in Table \ref{tab:gc_cost}. The number of AND gates represents the size of GC. The computational and communication costs include the total of both Cloud's and CSP's. GC's associated costs are linear to $n$ and bit precision $b$. By comparing Table \ref{tab:cost_dist} and Table \ref{tab:gc_cost}, it is clear that the GC-component dominates the overall communication cost of our protocols.
  
 \begin{table}[h!] 	
	\vspace{-0.45cm}
 	\centering
 	\scriptsize
 	 \caption{Costs of the GC component: Computation (comp.) and Communication (Comm.)} \label{tab:gc_cost}
	       \vspace*{-\baselineskip}
	        \resizebox{0.35\columnwidth}{!}{%
 	 	\begin{tabular}{|c|c|c|c|c|c|c|c|c|c|}
 		\hline
 		{Dataset}& AND Gates &Comp.(m) &Comm.(MB) \\
 		\hline
		ionosphere & 2,016,846 & 5.1 & 43.1 \\
 		credit &  8,840,000 &20.3 & 371.2  \\	
 		spambase  &  37,268,100& 47.2 & 1,202.6 \\
 		epileptic & 87,549,500 & 101.3 & 5,009.6\\
		synthetic & 695,400,000 &927.4 & 39791.1 \\
 		\hline
 	\end{tabular}
	}
	\vspace{-0.35cm}
 \end{table}

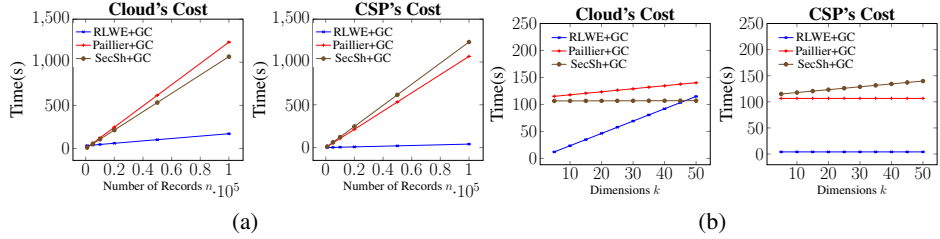
\begin{figure}[h]
\centering
\vspace{-0.55cm}
\begin{tabular}{c@{\thinspace}c}
\begin{tikzpicture}[scale=0.33]
 	\pgfplotsset{every axis legend/.append style={font=\large},every node near coord/.append style={font=\large}}
 	\begin{axis}
 	[xlabel={Number of Records $n$}, xlabel style={font=\Large},
	ymax=1500,
	title style={at={(0.5,1.1)},anchor=north,font =\huge},
	title ={\textbf{Cloud's Cost}},
 	ylabel={Time(s)}, ylabel style = {font = \huge},
	legend columns=1,legend style={at={(0.25,0.66)},anchor=south, font=\Large,draw=none},
	x tick label style = {font = \huge},
	y tick label style = {font = \huge}]
 	\addplot+[mark=x,error bars/.cd,
 	x dir=both,
 	y dir=both,y explicit]
 	table[x=N,y=Cost,col sep=tab]
 	{./data/rlwe_sp_growth_n.csv};
 	\addlegendentry{RLWE+GC}
	\addplot+[mark=+,error bars/.cd,
 	x dir=both
 	,y dir=both,y explicit]
 	table[x=N,y=Cost,col sep=tab]
 	{./data/paillier_sp_growth_n.csv};
	 \addlegendentry{Paillier+GC}	
	\addplot+[mark=oplus,error bars/.cd,
 	x dir=both
 	,y dir=both,y explicit]
 	table[x=N,y=Cost,col sep=tab]
 	{./data/secsh_sp_growth_n.csv};
	 \addlegendentry{SecSh+GC}	
 	\end{axis}
 	\end{tikzpicture}
\begin{tikzpicture}[scale=0.33]
 	\pgfplotsset{every axis legend/.append style={font=\large},every node near coord/.append style={font=\large}}
 	\begin{axis}
 	[xlabel={Number of Records $n$},ymax=1500, 
	xlabel style = {font = \Large},
	title style={at={(0.5,1.1)},anchor=north,font =\huge},
	title ={\textbf{CSP's Cost}},
 	ylabel={Time(s)},ylabel style = {font=\huge, yshift=-0.1cm},
	legend columns=1,legend style={at={(0.25,0.66)},anchor=south, font=\Large,draw=none},
	x tick label style = {font=\huge},
	y tick label style = {font = \huge}]
 	\addplot+[mark=x,error bars/.cd,
 	x dir=both,
 	y dir=both,y explicit]
 	table[x=N,y=Cost,col sep=tab]
 	{./data/rlwe_csp_growth_n.csv};
 	\addlegendentry{RLWE+GC}
	\addplot+[mark=+,error bars/.cd,
 	x dir=both
 	,y dir=both,y explicit]
 	table[x=N,y=Cost,col sep=tab]
 	{./data/paillier_csp_growth_n.csv};
	 \addlegendentry{Paillier+GC}		
	\addplot+[mark=oplus,error bars/.cd,
 	x dir=both
 	,y dir=both,y explicit]
 	table[x=N,y=Cost,col sep=tab]
 	{./data/secsh_csp_growth_n.csv};
	 \addlegendentry{SecSh+GC}		
 	\end{axis}
 	\end{tikzpicture}&
\begin{tikzpicture}[scale=0.33]
 	\pgfplotsset{every axis legend/.append style={font=\large},every node near coord/.append style={font=\large}}
 	\begin{axis}
 	[xlabel={Dimensions $k$},xlabel style = {font=\Large},ymax=250,
	title style={at={(0.5,1.1)},anchor=north,font =\huge},
	title ={\textbf{Cloud's Cost}},
 	ylabel={Time(s)}, ylabel style = {font=\huge},
	legend columns=1,legend style={at={(0.30,0.65)},anchor=south, font=\Large,draw=none},
	x tick label style = {font=\huge},
	y tick label style = {font=\huge}]
 	\addplot+[mark=x,error bars/.cd,
 	x dir=both,
 	y dir=both,y explicit]
 	table[x=Dim,y=Cost,col sep=tab]
 	{./data/rlwe_sp_growth.csv};
 	\addlegendentry{RLWE+GC}
	\addplot+[mark=+,error bars/.cd,
 	x dir=both,y dir=both,y explicit]
 	table[x=Dim,y=Cost,col sep=tab]
 	{./data/paillier_sp_growth.csv};
	 \addlegendentry{Paillier+GC}	
	 \addplot+[mark=oplus,error bars/.cd,
 	x dir=both,y dir=both,y explicit]
 	table[x=Dim,y=Cost,col sep=tab]
 	{./data/secsh_sp_growth.csv};
	 \addlegendentry{SecSh+GC}	
 	\end{axis}
 	\end{tikzpicture}
\begin{tikzpicture}[scale=0.33]
 	\pgfplotsset{every axis legend/.append style={font=\large},every node near coord/.append style={font=\large}}
 	\begin{axis}
 	[xlabel={Dimensions $k$},
	xlabel style = {font=\Large},
	ymax=250,
	title style={at={(0.5,1.1)},anchor=north,font =\huge},
	title ={\textbf{CSP's Cost}},
 	ylabel={Time(s)},ylabel style= {font=\huge,yshift={-0.1cm}},
	legend columns=1,legend style={at={(0.30,0.65)},anchor=south, font=\Large,draw=none},
	x tick label style = {font = \huge},
	y tick label style = {font = \huge}]
 	\addplot+[mark=x,error bars/.cd,
 	x dir=both,
 	y dir=both,y explicit]
 	table[x=Dim,y=Cost,col sep=tab]
 	{./data/rlwe_csp_growth.csv};
 	\addlegendentry{RLWE+GC}
	 \addplot+[mark=+,error bars/.cd,
 	x dir=both,y dir=both,y explicit]
 	table[x=Dim,y=Cost,col sep=tab]
 	{./data/paillier_csp_growth.csv};
	 \addlegendentry{Paillier+GC}		 
	 \addplot+[mark=oplus,error bars/.cd,
 	x dir=both
 	,y dir=both,y explicit]
 	table[x=Dim,y=Cost,col sep=tab]
 	{./data/secsh_csp_growth.csv};
	 \addlegendentry{SecSh+GC}			
 	\end{axis}
 \end{tikzpicture}\\
 \small (a) & \small (b)
 \end{tabular}
	\vspace{-0.25cm}	
 	\caption{Computation cost. (a) Over increasing records($n$) with fixed number of dimensions ($k = 20$). (b)  Over increasing dimensions($k$) (bottom) and fixed number of records ($n = 10,000$).}
 	\label{fig:cost_growth_n}
	\vspace{-0.45cm}
\end{figure}

 \textbf{Cloud and CSP Cost Scaling}
Now, we try to understand the relationship between the size of training data and associated costs using synthetic datasets of several sizes and dimensions. First, we fix the number of dimensions $k=20$ and see how number of records $n$ affects the costs. Figure \ref{fig:cost_growth_n} (a) shows that both Cloud's and CSP's costs in RLWE+GC grow much slower than the other two's. CSP's growth rates are almost same for SecSh+GC and Paillier+GC, as they involve the same number of decryption operations.

Figure \ref{fig:cost_growth_n} (b) depicts the effect of increasing the dimensions while fixing the number of records to  $n=10,000$. We observe that RLWE+GC cost for Cloud grows much faster for the larger dimensions. This is due to the associated dimension-wise RLWE replication cost in the matrix-vector multiplication. On the other hand, CSP's cost when using RLWE+GC is much lower than with the other two constructions, as the RLWE decryptions are much cheaper than that of Paillier. Both Cloud's and CSP's costs when using Paillier+GC and SecSh+GC stay almost flat as only $n$ dominates the overall cost.

\vspace{-0.35cm}
\subsection{Comparing with Other Methods}\label{subsec:tradeoff}
\vspace{-0.15cm}
In this section, we compare SecureBoost with the recently developed SecureML method \cite{mohassel17}. It implements the stochastic gradient-descent (SGD) learning based on secret sharing \cite{demmler15}, which is then used for logistic regression (LR) and neural network (NN) \cite{hastie01}. We tried different shapes of inner hidden layers and found the minimum-cost setting for satisfactorily handle the non-linearly separable synthetic dataset. SGD is conducted with a mini-batch size of 128 records in training. Both algorithms are run enough iterations until convergence.

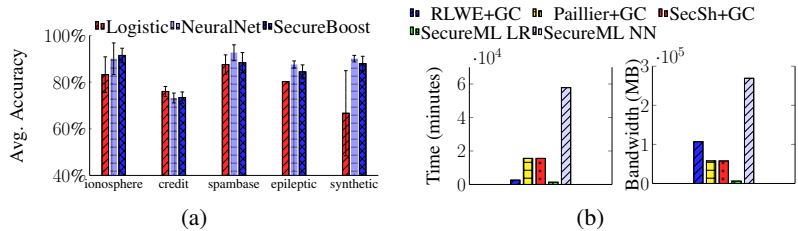
\begin{figure}[h]
\vspace{-0.65cm}
\centering
\begin{tabular}{c@{\thinspace}c}
\centering
 	\begin{tikzpicture}[scale=0.35]
 	\begin{axis}[enlarge x limits=-10,
 	ybar,x=65,
	width=8cm,
 	bar width=0.25cm, tick align=inside,
 	ymin=0, ymax=100,
 	axis x line*=bottom,axis y line*=left,
 	y axis line style={opacity=50},
	ymin=40,
 	tickwidth=2pt,
 	enlarge x limits=true,
 	legend columns=4,legend style={at={(0.5,0.98)},anchor=south,
 		font=\huge, draw=none},
 	ylabel={Avg. Accuracy},
 	ylabel style={font=\huge,yshift=0.4cm},
 	symbolic x coords={ionosphere,credit,spambase,epileptic,synthetic},
 	xtick=data,
	x tick label style={font=\Large},	
 	yticklabel={\pgfmathparse{\tick}\pgfmathprintnumber{\pgfmathresult}\%},
	yticklabel style ={font=\huge}
 	]
 	 \addplot[error bars/.cd, y dir=both,y explicit][draw=black, fill=red,fill opacity=0.85,postaction={pattern=north east lines}] coordinates {
 		(ionosphere,83.20)+-(7.65,7.65)
 		(credit,76.00)+-(2.16,2.16) 
 		(spambase,87.56)+-(4.13,4.13)
		(epileptic,80.27)+-(0.00,0.00)
		(synthetic,66.66)+-(18.25,18.25)};	
	\addplot[error bars/.cd, y dir=both,y explicit][draw=none,fill=blue,fill opacity=0.4,postaction={pattern=grid}] coordinates {
 		(synthetic,90.07)+-(1.40,1.40)
		(epileptic,87.59)+-(1.51,1.51)
		(ionosphere,90.04)+-(6.78,6.78)
 		(credit,73.20)+-(2.13,2.13) 
 		(spambase,92.61)+-(3.40,3.40)};
 	\addplot[error bars/.cd, y dir=both,y explicit][draw=black, fill=blue,fill opacity=0.85,postaction={pattern= crosshatch}] coordinates {
 		(synthetic,87.91)+-(3.18,3.18)
		(epileptic,84.41)+-(2.99,2.99)
		(ionosphere,91.45)+-(3.10,3.10)
 		(credit,73.4)+-(2.40,2.40) 
 		(spambase,88.39)+-(4.3,4.30) };
 	\legend{Logistic,NeuralNet,SecureBoost}
 	\end{axis}
 	\end{tikzpicture}&
	\hspace{0.2cm}
	\begin{tikzpicture}[scale=0.35]
 	\begin{axis}[
 	ybar=-4pt,
	width=7cm,
 	bar width=10pt,
 	ymin=0, ymax=70000,enlarge x limits=2,
 	axis x line*=bottom,
 	y axis line style={opacity=42},
 	legend columns=3,legend style={at={(0.8,1.2)},draw=none,anchor=south,
 		font=\huge},
 	ylabel={Time (minutes)},
 	ylabel style={font=\huge}, ylabel style ={yshift=0.3 cm},
	yticklabel style={font=\huge},
 	symbolic x coords={RLWE,Paillier,SecShare,SecMLLR,SecMLNN},
	hide obscured x ticks=true,
 	xtick=\empty,
 	yticklabel={\pgfmathparse{\tick}\pgfmathprintnumber{\pgfmathresult}}]
 	 \addplot[error bars/.cd, y dir=both,y explicit][draw=black, fill=blue,fill opacity=0.85,postaction={pattern=north east lines}] coordinates {
		 (RLWE,2590)};
	 \addplot[error bars/.cd, y dir=both,y explicit][draw=black, fill=yellow,fill opacity=0.85,postaction={pattern=grid}] coordinates {
		(Paillier,15571)};
	 \addplot[error bars/.cd, y dir=both,y explicit][draw=black, fill=red,fill opacity=0.85,postaction={pattern=dots}] coordinates {
		 (SecShare,15570)};
	\addplot[error bars/.cd, y dir=both,y explicit][draw=black, fill=green,fill opacity=0.85,postaction={pattern=north west lines}] coordinates {
		(SecMLLR,1340.1)};
	 \addplot[error bars/.cd, y dir=both,y explicit][draw=black, fill=blue,fill opacity=0.15,postaction={pattern=north east lines}] coordinates {
		(SecMLNN,57832.5)};
	\legend{RLWE+GC, Paillier+GC,SecSh+GC, SecureML LR,SecureML NN} 
 	\end{axis}
	\end{tikzpicture}
	\hspace{-2.0cm}
\begin{tikzpicture}[scale=0.35]
	\begin{axis}[
 	ybar=-4pt, 
	width=7cm,
	enlarge x limits=2,
 	bar width=10pt,
 	ymin=0, ymax=300000,
 	axis x line*=bottom, ylabel near ticks, yticklabel pos=left, yticklabel style={font=\huge},
	 y axis line style={opacity=50},
 	legend columns=2,legend style={at={(0.5,0.99)},draw=none,anchor=south,
 		font=\Large},
 	ylabel={Bandwidth (MB)}, ylabel style ={yshift=-0.3 cm},
 	ylabel style={font=\huge},
 	symbolic x coords={RLWE,Paillier,SecShare,SecMLLR,SecMLNN},
	hide obscured x ticks=true,
	xtick=\empty,
	yticklabel={\pgfmathparse{\tick}\pgfmathprintnumber{\pgfmathresult}}]
 	  \addplot[error bars/.cd, y dir=both,y explicit][draw=black, fill=blue,fill opacity=0.85,postaction={pattern=north east lines}] coordinates {
	  	(RLWE,106891.1)};
	  \addplot[error bars/.cd, y dir=both,y explicit][draw=black, fill=yellow,fill opacity=0.85,postaction={pattern=grid}] coordinates {
		(Paillier,57662.2)};
	 \addplot[error bars/.cd, y dir=both,y explicit][draw=black, fill=red,fill opacity=0.85,postaction={pattern=dots}] coordinates {
		 (SecShare,57663.5)};
		\addplot[error bars/.cd, y dir=both,y explicit][draw=black, fill=green,fill opacity=0.85,postaction={pattern=north west lines}] coordinates {
		(SecMLLR,6054.57)};
	 \addplot[error bars/.cd, y dir=both,y explicit][draw=black, fill=blue,fill opacity=0.15,postaction={pattern=north east lines}] coordinates {
		(SecMLNN,269176.58)};
 	\end{axis}
 	\end{tikzpicture}\\
	\small (a) &\small (b)
	\end{tabular}
	\vspace{-0.3cm}
 	\caption{(a) Comparison of model accuracy: Secure-Boost vs. SecureML - Logistic Regression and Neural Network. (b) Overall cost comparison: SecureBoost constructions vs. SecureML neural network and SecureML logistic regression for the synthetic dataset.}	
 	\label{fig:comparison}
	\vspace{-0.55cm}
\end{figure}

Figure \ref{fig:comparison} (a) shows that SecureBoost and SecureML-NN perform similarly, while SecureML-LR due to its inherent linearity \cite{hastie01} underperforms significantly on the non-linearly separable data. This result can also be reproduced and verified with the scripts we have uploaded online \footnote{https://sites.google.com/site/testsboost/}.  Figure \ref{fig:comparison} (b) shows that SecureBoost constructions are more efficient than SecureML neural network. The cost patterns will vary for different datasets due to the varying number of training epochs. For this specific dataset, SecureBoost takes 200 iterations, while SecureML NN takes 20 epochs to converge. Logistic regression converges quickly within $10$ epochs but gets stuck on a non-optimal result. It appears the per-iteration cost of SecureML NN is much higher.   

 
\vspace{-0.25cm}
\subsection{Effect of Releasing $I_t$}
\vspace{-0.15cm}

\begin{figure}[h]
	\vspace{-0.45cm}
	\centering
	\includegraphics[width=0.23\linewidth]{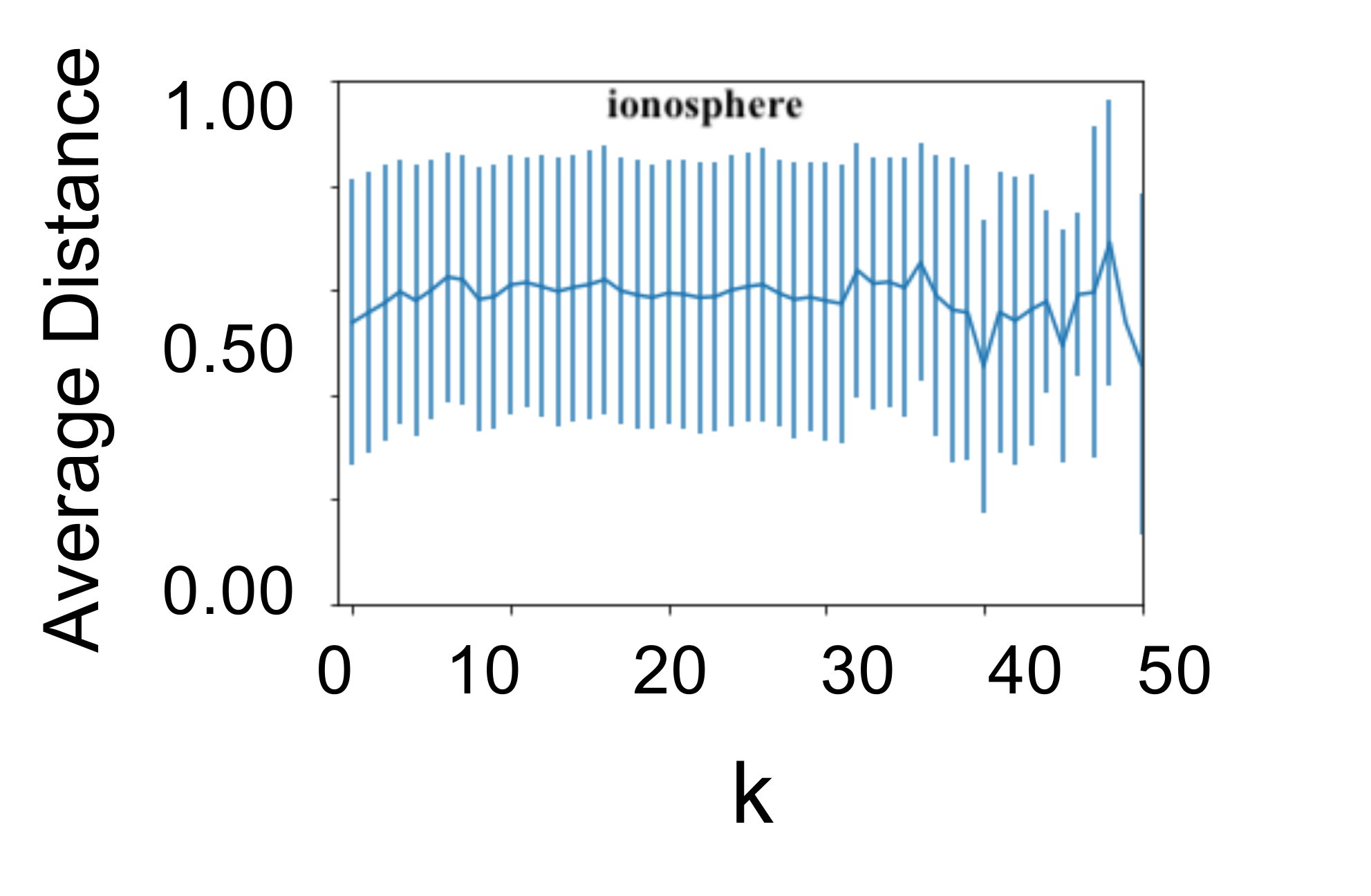}
	\includegraphics[width=0.23\linewidth]{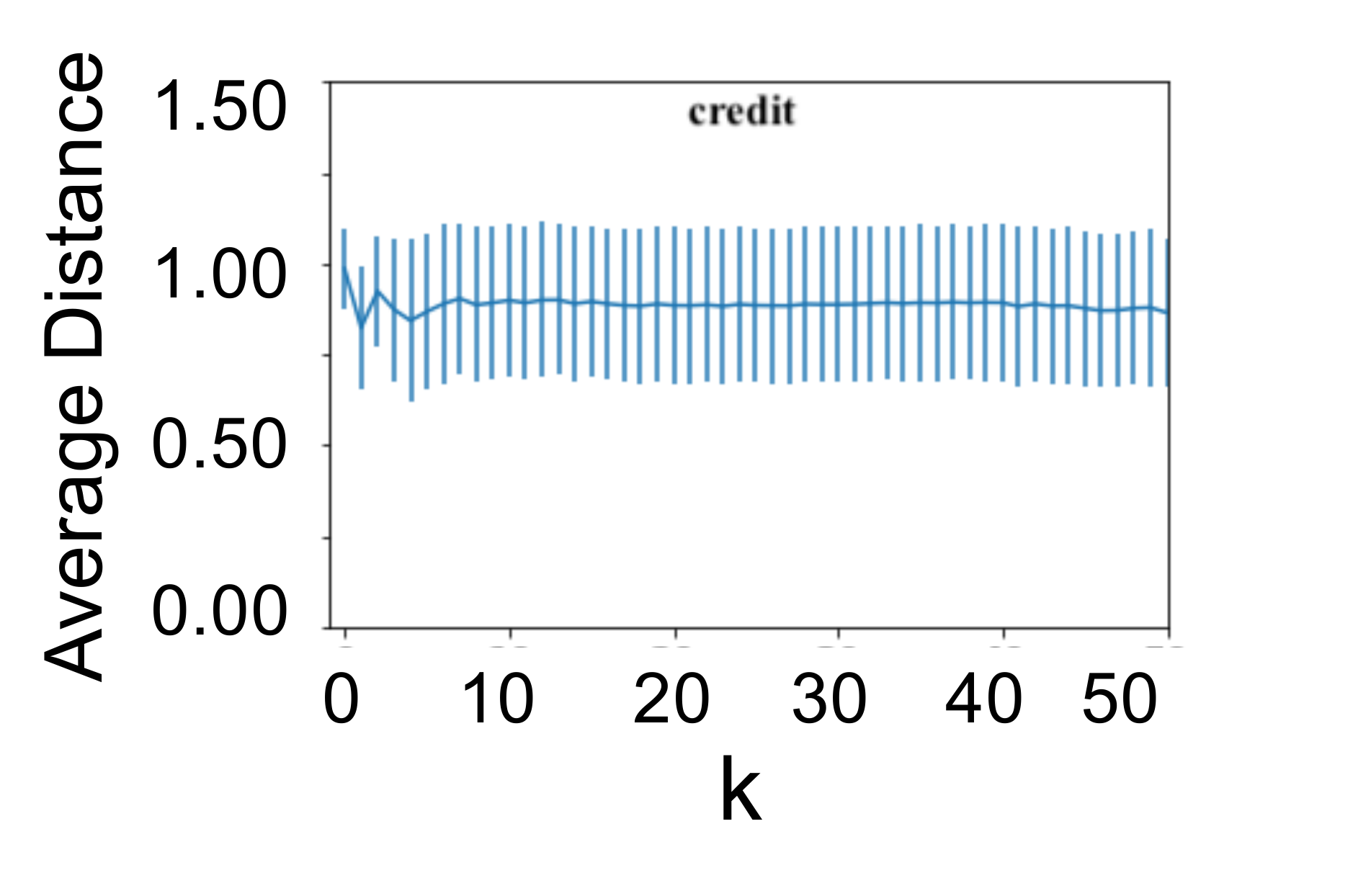}
	\includegraphics[width=0.23\linewidth]{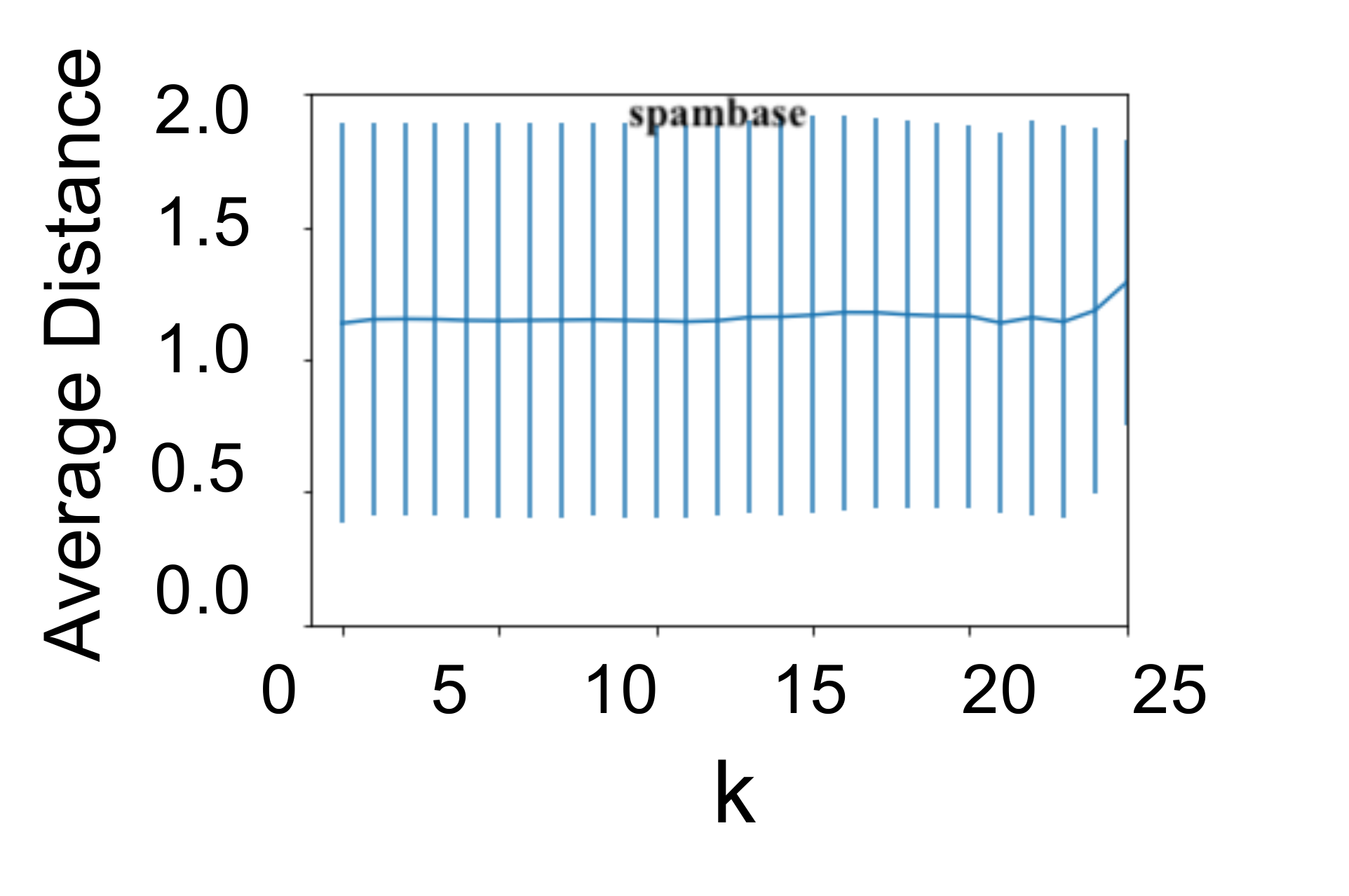}
	\includegraphics[width=0.23\linewidth]{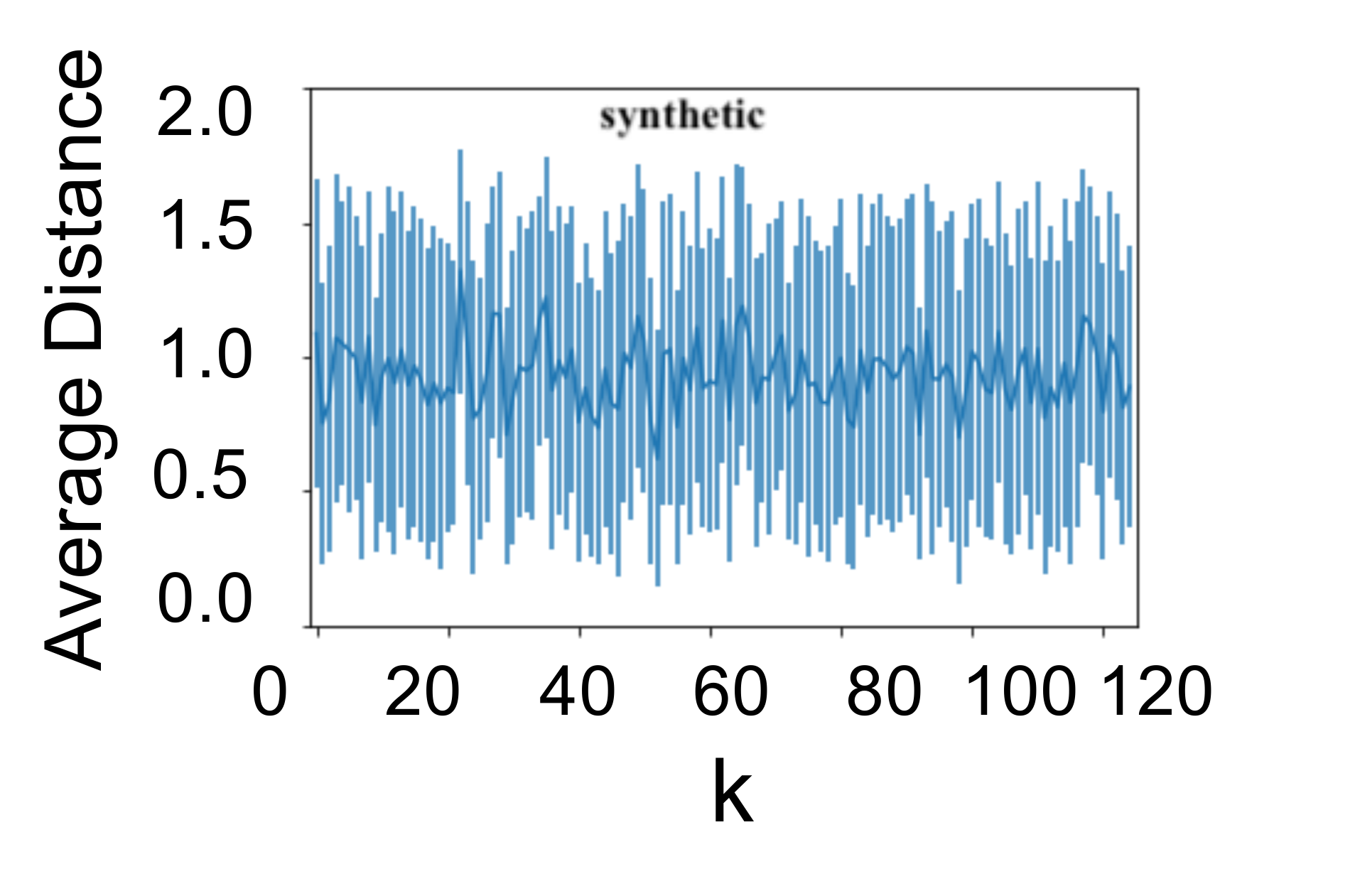}
	\vspace{-0.4cm}
	\caption{Avg. distance between record-pairs generating characterization vectors differing by $k$-bits.}
	\label{fig:char_dist}
	\vspace{-0.7cm}
\end{figure}

We want to verify if similar characterization vectors infer similar training records to understand the leaked information by $I_t$. Figure \ref{fig:char_dist} measures the average Euclidean distances between the training record pairs corresponding to the characteristic vectors differing by $k$ bits. It is evident that the similarity of characterization vectors does not infer the similarity of training records.

\section{Related Work}\label{sec:related_work}
\vspace{-0.35cm}
The current implementations of FHE are still too expensive to apply on complex functions. ML Confidential \cite{graepel12} shows that simple linear models can be learned by a semi-honest Cloud from FHE-encrypted data with acceptable costs. However, these simple models are unable to handle non-linearly separable datasets. Lu et al. \cite{lu16stat} show that PCA and linear regression can be implemented on FHE encrypted data with reasonable costs for a strictly small number of iterations in the algorithms. Moreover, the comparison operation based on FHE is very expensive \cite{lu16stat}, which hinders the FHE's application in many algorithms.

Despite new optimization of GC with techniques, such as free XOR gates \cite{kolesnikov08}, half AND gates \cite{zahur15}, and OT Extension \cite{asharov13}, its adaptation in confidential frameworks is still costly.  Nikolaenko et al. \cite{niko13,niko13sp} use FastGC \cite{huang11} and AHE to implement matrix factorization and linear ridge regression solutions. Use of  GCs in the expensive operations led these protocols to suffer from unbearable communication costs between CSP and Cloud. In our designs, we carefully craft the primitive operations to minimize the performance impact of the GC-related operations.

Demmler et al. \cite{demmler15} have shown that basic matrix operations can be implemented on random shares held by different parties when using secret sharing secure multi-party computations. SecureML \cite{mohassel17} utilized these operations and GC to implement the gradient-descent learning method with a two-server model. However, we note that these models are more expensive than ours to achieve the same level of model quality.

Users may also submit locally perturbed data that satisfy locally differential privacy (e.g., RAPPOR \cite{erlingsson14}). However, the model quality is significantly affected by the reduced data quality, and the models are also exposed to model-inversion attacks \cite{fredrikson14,shokri16}.

Gamb's et al.\cite{gambs07}  proposed algorithms enabling two or more participants to construct a boosting classifier, however, their goal is to train a combined model without sharing the horizontally partitioned training data with one another, not outsourcing it.  

\vspace{-0.45cm}
\section{Conclusion}\label{sec:conclusion}
\vspace{-0.25cm}
We develop the SecureBoost protocol for data owners to learn high-quality boosted classification models from encrypted or randomly partitioned users' data using public Cloud. The key idea is to use random linear classifiers as the base classifiers to simplify the protocol design. Two constructions: HE+GC and SecSh+GC have been developed, using a novel combination of homomorphic encryption, garbled circuits, and randomized secret sharing to protect the confidentiality and achieve efficiency. We formally analyze the security of the protocol and show that SecureBoost constructions satisfy the universally composable security for multiparty computation. Our experimental evaluation examines the intrinsic relationships among the primitive selection, cost distribution, and model quality. Our results show that the SecureBoost approach is very practical in learning high-quality classification models. Our constructions are the first batch of boosting protocols with practical costs, compared to the expenses of the start-of-the-art implementation of other major predictive modeling methods (e.g., Neural Networks by SecureML). We will extend the study to explore the effect of sub-sampling the training data and differentially private release of the leakage function in the future. Similarly, we will extend the work to multi-class classification problem and other types of boosting.

\appendix

\section{Appendix} \label{Appendix}

\subsection{Confidential Decision Stump Learning}\label{App:DS_boost}
\vspace{-0.15cm}
As there is no confidential DS learning algorithm reported, we present our initial design of DS learning that fits our boosting framework. Learning DS involves finding the optimal split for each feature in the training data with maximum information gain. The original algorithm takes $O(n\log n)$ comparisons to sort the values for each feature. However, sorting the dimensions may reveal the ordering information and breach data confidentiality, therefore, sorting may not be used in the confidential version of DS learning. Instead, we use a fixed binning scheme - i.e., partitioning the domain of each normalized dimension (e.g., (-4, 4)) into $s$ bins and enumerate all possible decision stumps -  for two-class problems and $k$ dimensions, there are $2sk$ such stumps (each split value gets two \emph{conjugate} stumps: e.g., Stump 1: if $X_j < v_j$ return 1 else return 0, Stump 2: if $X_j \geq v_j$ return 1 else return 0). We will describe the HE+GC construction for DS learning here. 

The users encrypt their records $E(x_i)$ and labels $E(y_i)$, with $y_i \in \{0,1\}$,  separately with the public key distributed by the CSP. (1) Cloud will start to evaluate each of the $sk$ decision stumps for every record with a slightly modified version of GC described in Section \ref{sec:core_alg_rlwe}. Specifically, for each instance $(x_i, y_i)$, it will securely check whether the class label $y_i$ matches the classifier output, e.g., if $X_j < v_j$ return 1 else return 0. Similarly, the evaluation of each DS will give an indicator vector $I_r$, $r=1..sk$, where 1 represents prediction error, reverse to the indicator vector described in Section \ref{sec:protocol}, $I_r$ is known to both Cloud and CSP. We can flip the indicator vector for the conjugate DS. (2) CSP starts a \emph{base classifier selection} process, and computes the weight $\alpha_t$ for each selected DS $h_t(x)$. Specifically, with training sample weights (initialized to $1/n$), $w_i$, at iteration $i$, CSP will find one of the $sk$ DSes that minimizes the weighted error,  $\arg\min_{r} \dot(I_r, w_i)$, for $r=1..sk$. In the end, CSP only knows the index of the DS. It does not know the base classifier parameters, i.e. neither $X_j$ nor $v_j$. Note that this step does not involve decryption and encryption. (3) The indices of the selected DSes and $\alpha_i $ are submitted by CSP to Data Owner. Data Owner can retrieve the actual DSes from Cloud.   

Therefore, the overall cost is dominated by the $sk$ rounds of evaluation in stage (1), not subject to the number of selected base classifiers. To get results close enough to the DS-based boosting model, we may need to take finely divided bins, e.g., s=100. For a 10-dimension dataset, the cost is about equivalent to trying 1000 base classifiers in the RLC protocol.  Furthermore, CSP takes a significant amount of storage and computing burden --- it will need to keep all the $sk$ indicator vectors for DS selection, the size of which is much larger than the original data, and conduct $sk\tau$ dot products on plaintext if the final model contains $\tau$ base classifiers.

\vspace{-0.35cm}

\bibliographystyle{abbrv}
\bibliography{./paper}


\vspace{-0.2cm}

\end{document}